\documentclass[12pt]{article}
\usepackage[utf8]{inputenc}
\usepackage{graphicx}
\usepackage{natbib}
\usepackage{color}
\usepackage{amsmath}
\usepackage{amsfonts}
\usepackage{hyperref}
\usepackage[algo2e]{algorithm2e} 
\usepackage{algorithm}
\usepackage{verbatim}
\usepackage{caption}
\usepackage{setspace}
\usepackage{subcaption}
\usepackage{multirow}
\usepackage[left=1in,right=1in,top=1in,bottom=1in]{geometry}
\newcommand{\bc}{\boldsymbol c}

\newcommand{\bA}{{\bf A}}
\newcommand{\bC}{{\bf C}}
\newcommand{\bN}{{\bf N}}
\newcommand{\bZ}{{\bf Z}}

\title{Graph sub-sampling for divide-and-conquer algorithms in large networks}
\author{Eric Yanchenko \\ Akita International University\\
eyanchenko@aiu.ac.jp}

\begin{document}

\maketitle
\thispagestyle{empty}

%\doublespacing

\begin{abstract}
\noindent
As networks continue to increase in size, current methods must be capable of handling large numbers of nodes and edges in order to be practically relevant. Instead of working directly with the entire (large) network, analyzing sub-networks has become a popular approach. Due to a network's inherent inter-connectedness, however, sub-sampling is not a trivial task. While this problem has gained popularity in recent years, it has not received sufficient attention from the statistics community. In this work, we provide a thorough comparison of seven graph sub-sampling algorithms by applying them to divide-and-conquer algorithms for community structure and core-periphery (CP) structure. After discussing the various algorithms and sub-sampling routines, we derive theoretical results for the mis-classification rate of the divide-and-conquer algorithm for CP structure under various sub-sampling schemes. We then perform extensive experiments on both simulated and real-world data to compare the various methods. For the community detection task, we found that sampling nodes uniformly at random yields the best performance, but that sometimes the base algorithm applied to the entire network yields better results both in terms of identification and computational time. For CP structure on the other hand, there was no single winner, but algorithms which sampled core nodes at a higher rate consistently outperformed other sampling routines, e.g., random edge sampling and random walk sampling. Unlike community detection, the CP divide-and-conquer algorithm tends to yield better identification results while also being faster than the base algorithm. The varying performance of the sampling algorithms on different tasks demonstrates the importance of carefully selecting a sub-sampling routine for the specific application.

\end{abstract}

\noindent
{\it Keywords:} Exploration sampling; Community detection; Core-periphery; Mis-classification rate; Network analysis

\section{Introduction}\label{sec:intro}

Graphs, or networks\footnote{We will use these terms interchangeably in this work.}, provide a simple and intuitive model for interconnected systems where objects or entities are represented as nodes, and their relationships are denoted as edges. Networks have been applied to many real-world applications, ranging from friendships \citep{girvan2002community} to brain connectivity \citep{telesford2011brain}. As their popularity has grown, so too has their size: it is not uncommon for researchers to now work with thousands, millions or even billions of nodes and edges \citep[e.g.,][]{backstrom2006, rozemberczki2019}. Community detection \citep{mukherjee2021two, fang2024online, cardoso2024greedy, chakrabarty2025subsampling}, hypothesis testing \citep{ghoshdastidar2018practical,chakraborty2024scalable} and link prediction \citep{zhang2018link, kumar2020link} are just a few examples where researchers must handle large graphs. Thus, modern network analysis methods must scale to such datasets if they are to be practically useful.

One popular approach for working with large data sets in general, and large graphs in particular, is sub-sampling. The idea is simple: first the researcher divides the data into smaller sub-sets before performing analyses on the sub-sets. The analyses are then combined to give a result on the original (larger) data set. Generally, performing an analysis multiple times on smaller data sets is much faster than applying it once to the entire dataset. While sub-sampling and divide-and-conquer approaches have found great success across various statistical, machine learning and computer science tasks \citep{jordan2012divide}, there are some unique challenges to applying these techniques to graph-valued data. The main difficulty is that the nodes and edges which make up the graph are, by their very nature, dependent. As a graph can have a highly complex topology, it is unclear if these features will be preserved in the sub-graphs, and if not, it is unclear how this affects the analysis. Thus, it is paramount to understand the effect of the graph sub-sampling routine on the task of interest.

Despite its importance, graph sampling is still an open-problem, and, in particular, has received less attention from the statistics community. \cite{leskovec2006sampling} study the similarity between various measures computed on the original graph and their sub-graphs, including: clustering coefficients, degree distributions, largest eigenvectors and more. The authors find that while edge sampling performs poorly and random node sampling performs surprisingly well, random walk and forest fire samplers yield sub-networks most similar to the original network. Instead of comparing different sub-sampling strategies, other works focus their study on a single paradigm, such as random walks \citep[e.g.,][]{ribeiro2012sampling, li2015random, chiericetti2016sampling}

Several recent works have proposed novel graph sampling algorithms for various tasks. \cite{wang2023adaptive} develop an adaptive edge sub-sampling algorithm and apply it to find degree, clustering coefficient and $K$-core distributions. Similarly, \cite{chen2024extracting} propose an edge-reinforced random walk sub-sampling strategy and show that sub-graphs extracted using the algorithm are similar to the original network, while \cite{wang2024supports} propose a motif-based sub-sampler for support estimation. 

Recently, \cite{littleballoffur} develop a package to easily implement various sub-sampling routines. The new package is then applied to a variety of tasks, from computing average degrees to node classification. \cite{rovzanec2024go} provide a similar package implementing various sub-sampling algorithms in the Go programming language. To our knowledge, there are not any works studying the effects of graph sub-sampling from a statistical angle, nor specifically studying their effect on divide-and-conquer algorithms.

In this work, we seek to close this gap by comparing graph sub-sampling methods on divide-and-conquer algorithms. We focus on two of the most important meso-scale structures: community structure \citep{newman2006modularity}, where nodes cluster into highly connected groups, and core-periphery structure \citep{yanchenko2023core}, where nodes are either in the densely-connected core or sparsely-connected periphery. Our main contribution is a thorough comparison, both theoretically and empirically, of seven sub-sampling routines on the performance of divide-and-conquer algorithms for identifying these structures.
After introducing the methods, we recap the theoretical results from \cite{mukherjee2021two} before deriving novel theoretical results for the CP method of \cite{yanchenko2022divide}. In particular, we derive the general mis-classification rate of the algorithm and give specific results for several sub-sampling methods. Finally, we apply these methods to various synthetic and real-world networks. We find that random node and random node neighbor yield the best community structure identification results, while routines which sample core nodes with a high probability, e.g., edge sampling and random walk, perform the best for the CP identification task. Indeed, the sub-sampling routines that perform well for one task, generally did not perform as well on the other, underscoring the importance of carefully choosing the sub-sampling algorithm for the specific problem. Moreover, the performance of the base algorithm relative to the divide-and-conquer approach also differed depending on the task. Notably, the CP divide-and-conquer algorithm outperformed the base algorithm both in terms of detection ability and computation time.

The layout of the rest of the paper is as follows. In Section \ref{sec:method} we introduce the divide-and-conquer algorithms as well as the various sub-sampling routines. Section \ref{sec:theory} is devoted to theoretical results while Sections \ref{sec:sim} and \ref{sec:real} apply the methods to simulated and real-world networks, respectively. Finally, we share concluding thoughts in Section \ref{sec:conc}.

\section{Methodology}\label{sec:method}
We begin by discussing the divide-and-conquer algorithms for community and CP structure, followed by presenting seven graph sub-sampling algorithms.

\subsection{Notation}
Let $G=(V,E)$ be a graph with $|V|=n$ nodes and $|E|=m$ edges. We also define $\bA$ as the $n\times n$ adjacency matrix corresponding to graph $G$ such that $A_{ij}=1$ if nodes $i$ and $j$ are connected by and edge, and 0 otherwise.\footnote{In this work, we only consider undirected, unweighted networks without self-loops, but the ideas can easily be generalized.} Let $K$ be the known (integer) number of communities, and let $\bZ\in\{0,1\}^{n\times K}$ correspond to the true community labels where $Z_{ik}=1$ if node $i$ is in community $k$ for $k=1,\dots,K$. We define $\bC=\bZ\bZ^T\in\{0,1\}^{n\times n}$ as the clustering matrix where $C_{ij}=1$ if nodes $i$ and $j$ are in the same community, and 0 otherwise. Notice that $\bC$ does not depend on the exact community labels (which are equivalent up to permutation) but only on the pair-wise relationship between nodes. 

\subsection{Divide-and-conquer for community structure}
\cite{mukherjee2021two} propose $\mathsf{PACE}$, a divide-and-conquer algorithm to identify community structure in networks. The algorithm randomly samples sub-graphs, and then applies a community detection algorithm to each sub-graph. This step repeats many times before the more challenging step of ``stitching'' together the results from the sub-samples. Since community labels are not uniquely identifiable, this combining step is non-trivial. To avoid this problem, the authors estimate the clustering matrix $\bC$ instead of directly estimating the community labels as the clustering matrix is unique. This clustering matrix can then be used to extract the community labels. 

$\mathsf{PACE}$ begins by randomly sampling a sub-graph $S^{(b)}$ of graph $G$ with $qn$ nodes where $q\in(0,1)$ is the proportion of nodes in the sub-sample.\footnote{Note that $q$ may be chosen to ensure $qn$ is an integer, or $qn$ can simply be rounded to the nearest integer.} With a sub-sample in hand, we can apply any community detection algorithm to this sub-sample, e.g., spectral \citep{ng2002spectral}, modularity maximization \citep{girvan2002community}, semi-definite programming \citep{cai15}, etc. The resulting membership vectors of these clustering algorithms allow us to construct $\hat \bC^{(b)}\in\{0,1\}^{n\times n}$ where $\hat C^{(b)}_{ij}=1$ if both nodes $i$ and $j$ appear in sub-graph $S^{(b)}$ and were assigned to the same community, and 0 otherwise. We repeat this process for $B$ sub-graphs, yielding $\hat \bC^{(1)},\dots, \hat \bC^{(B)}$.

The key step in this algorithm is combining the results from the $B$ sub-graphs. Let $N_{ij}$ be the number of times that both nodes $i$ and $j$ were chosen in the same sub-sample. Then the combined estimator $\hat \bC\in[0,1]^{n\times n}$ of the clustering matrix is defined by
\begin{equation}\label{eq:pace}
    \hat C_{ij}
    =\mathbb I(N_{ij}>\beta) \frac{\sum_{b=1}^B \hat C_{ij}^{(b)}}{N_{ij}}
\end{equation}
where $\mathbb I(\cdot)$ is the indicator function, and $1\leq \beta\leq B$ is a tuning parameter. The combined estimator $\hat \bC$ is the proportion of sub-samples where nodes $i$ and $j$ were both sampled and assigned to the same community. We only compute this quantity, however, for node pairs which were sampled a sufficient number of times ($>\beta$ times), otherwise the estimate is set to 0. Lastly, we can apply a clustering algorithm, e.g., $k$-means, to $\hat \bC$ in order to obtain the final community membership vector. The full algorithm is reported in Algorithm 1.

\begin{algorithm}[]
\SetAlgoLined
\KwResult{Community membership vector $\hat{\bc}\in\{1,\dots,K\}^n$}
 {\bf Input: }Graph $G$, number of communities $K\in\mathbb Z^+$, proportion of nodes to sub-sample $q\in(0,1)$, number of sub-samples $B\in \mathbb Z^+$, sub-sampling scheme $\mathcal S$, clustering algorithm $\mathcal A$, tuning parameter $\beta\in\{1,2,\dots,B\}$\;

  \For{$B$ times}{
  Randomly sample $G$ using $\mathcal S$, resulting in sub-network $S^{(b)}$\;

  Apply $\mathcal A$ to $S^{(b)}$\;
  
  Obtain clustering matrix $\hat \bC^{(b)}\in\{0,1\}^{qn\times qn}$\;
}

Compute $\bN$ where $N_{ij}$ is the number of times nodes $i$ and $j$ were sampled together in a sub-graph.\;

Compute $\hat \bC$ where $\hat C_{ij}=\mathbb I(N_{ij}>\beta)\displaystyle\frac{\sum_{b=1}^B \hat C_{ij}^{(b)}}{N_{ij}}$\; 

Perform $k$-means clustering on $\hat \bC$ with $K$ clusters to obtain $\hat{\bc}\in\{1,\dots,K\}^n$\;
 
\caption{$\mathsf{PACE}$ for community structure}
\end{algorithm}

The key advantage of this algorithm is that it estimates $\bC$, the clustering matrix, instead of directly estimating the community memberships $\bZ$, as the clustering matrix is independent of label permutations. Moreover, the estimate of $\hat C_{ij}$ is intuitive, namely the number of times nodes $i$ and $j$ were assigned to the same community divided by the number of times they were sampled together, given they were sampled together a sufficient number of times. Checking if $N_{ij}>\beta$ acts as a smoothing step by discarding the estimates for node pairs which only appeared in a few sub-graphs. Of course, $\beta$ is a tuning parameter that must be chosen by the user. As suggested by the authors of $\mathsf{PACE}$, we set $\beta$ to be the 40th percentile of $N_{ij}$. Additionally, note that the final $k$-means clustering step could fail by providing zero- or one-node clusters, which is a limitation of this method.

The main focus of this paper is the sub-sampling step of the divide-and-conquer algorithm, a topic that the authors of $\mathsf{PACE}$ give some attention too (see Section 2.3 of \cite{mukherjee2021two}). In particular, as possible sub-sampling schemes, they suggest: randomly sampling nodes; randomly sampling a node and including all of its neighbors within distance $h$ for some integer $h$ ($h$-hop neighborhoods); $1$-hop neighborhoods but removing the root node (ego neighborhoods); randomly sampling a node, including all of its neighbors and all of the neighbors of the neighbors, etc. (onion neighborhoods); or sampling sub-graphs with roots at high degree nodes. For most experiments, the authors employ random node sampling.

\subsection{Divide-and-conquer for core-periphery structure}
We now turn our attention to the divide-and-conquer algorithm for detecting core-periphery (CP) structures in large networks proposed by \cite{yanchenko2022divide}.

\subsubsection{Core-periphery metric}
Before presenting the divide-and-conquer algorithm, we first discuss the metric from \cite{BORGATTI2000} (BE) used to quantify the CP structure of a network. The authors define the core as a set of nodes which are highly connected with each other, as well as with the peripheral nodes. In particular, they find the correlation between the observed network and a network with ``ideal'' CP structure. Let $\bA$ again be the adjacency matrix for the observed network and $\bc\in\{0,1\}^n$ be some CP labels where $c_i=1$ if node is assigned to the core, and 0 otherwise. Then the BE metric is
\begin{equation}\label{eq:BE}
    \rho(\bA,\bc)
    =\mathsf{Cor}(\bA,\Delta_{\bc})
\end{equation}
where $\mathsf{Cor}(\bA,{\bf B})$ is the Pearson correlation of the vectorized upper triangular parts of matrices $\bA$ and ${\bf B}$, and $\Delta_{\bc}$ represents the ideal CP structure with $(\Delta_{\bc})_{ij}=c_i+c_j-c_ic_j$. In words, $(\Delta_{\bc})_{ij}=1$ if either node $i$ or $j$ is in the core, and 0 otherwise. The metric in \eqref{eq:BE} can be optimized using the greedy algorithm proposed in \cite{yanchenko2022divide} to approximate
\begin{equation}\label{eq:opt}
    \hat{\bc}
    =\arg\max_{\bc\in\{0,1\}^n}\mathsf{Cor}(\bA,\Delta_{\bc}).
\end{equation}

\subsubsection{Divide-and-conquer algorithm}
With a metric to quantify the CP structure of a network, we can present the divide-and-conquer algorithm. We slightly modify the algorithm in \cite{yanchenko2022divide} for reasons that will be explained later. The idea is to find the CP structure on small sub-networks of the network and then combine these results together to yield the CP labels on the entire network. Specifically, let $G$ be the graph with $n$ nodes, $q\in(0,1)$ be the proportion of nodes to sub-sample and $B$ the number of sub-sample to draw. First, we sample a sub-graph with $qn$ nodes using some graph sub-sampling routine, and find the optimal CP labels from \eqref{eq:opt}. We repeat this process $B$ times and output $\hat{\bc}\in[0,1]^n$ where $\hat c_i$ is the number of times node $i$ was assigned to the core divided by the total number of samples $B$. The full algorithm is presented in Algorithm 2.

\begin{algorithm}[]
\SetAlgoLined
\KwResult{Core-periphery proportions $\hat{\bc}$}
 {\bf Input: }Graph $G$, proportion of nodes to sub-sample $q\in(0,1)$, number of sub-samples $B\in\mathbb Z^+$, sub-sampling scheme $\mathcal S$\;
 
  \For{$B$ times}{
  Randomly sample $G$ using $\mathcal S$ to obtain sub-network $S^{(b)}$\;

  Obtain CP labels of $S^{(b)}$, $\hat{\bc}^{(b)}\in\{0,1\}^{qn}$\;}
  
  Obtain $\hat\bc$ as $\hat c_i=\displaystyle\frac{1}{B}\sum_{b=1}^B\hat c_i^{(b)}$\;
 
\caption{Divide-and-conquer for core-periphery structure}
\end{algorithm}

This algorithm is slightly different from the original paper. Originally, the output was the number of times node $i$ was assigned to the core divided by the number of times that node $i$ was sampled. Indeed, this approach more closely resembles the stitching routine of the $\mathsf{PACE}$ algorithm. The reason that $\mathsf{PACE}$ divides by the number of times that nodes $i$ and $j$ were sub-sampled together, $N_{ij}$, and then only keeps the estimate if $N_{ij}>\beta$ is that the estimate $\hat C_{ij}$ would be unreliable if it was constructed from only a few sub-samples. For the CP task, however, if a node is sub-sampled infrequently, this gives us {\it more} information, rather than less. Indeed, core nodes are those that are more central or influential in the network. Therefore, if a node is sampled infrequently, then it is unlikely to be important and also unlikely to be a part of the core. Thus, dividing by the total number of samples will decrease the estimate of $\hat c_i$ for infrequently sampled nodes. Implicitly, this means we should favor sub-sampling methods which have a high probability of sampling core nodes. As we will see, this small change not only makes proving theoretical statements easier, but also improves the empirical performance of the algorithm.

\subsection{Sub-graph sampling}\label{sec:sub}
Finally, we discuss seven sub-sampling algorithms. We select a variety of sampling schemes including: node-based (random and proportional to degree); edge-based (random) and exploration (breadth-first search, depth-first search, random node-neighbor and random walk).

\paragraph{Random node (RN):} The simplest method is to randomly sample nodes without replacement. Each node has an equal probability of being selected, and once all $qn$ nodes are chosen, then the edges between these nodes make up the sub-graph. This method was originally considered in both \cite{mukherjee2021two} and \cite{yanchenko2022divide}. While this is the simplest sampling scheme, it is unclear if it will preserve important properties of the network. For example, \cite{richardson2003trust} show that if the degrees of the full network have a power-law distribution, then RN sampling will not necessarily yield the same degree distribution on the sub-graphs.

\paragraph{Degree node (DN):} In this method, nodes are again sampled at random, but now their probability of being sampled is proportional to their degree. Once the $qn$ nodes are sampled, then the sub-graph is completed with all associated edges between sampled nodes. Mathematically, this means sampling without replacement from a distribution with probability mass function $\mathsf{P}(Z=j)=\displaystyle\frac{d_j}{\sum_{k=1}^n d_k}$ where $d_j$ is the degree of node $j\in\{1,2,\dots,n\}$. As \cite{leskovec2006sampling} point out, this approach will likely lead to sub-graphs which have a different degree distribution from the original graph, while also being denser.

\paragraph{Random edge (RE):} Instead of randomly sampling nodes, this approach samples edges uniformly at random. Edges are drawn with equal probability and the two nodes connected by the sampled edge are added to the sub-graph. This process continues until $qn$ nodes have been sampled, at which point the remaining edges between sampled nodes are also added. Adding the remaining edges between sampled nodes is sometimes called the Induction Step \citep{ahmed2013network} and will be used in all remaining algorithms as well. While edges can be selected with equal probability, if the graph has edge weights, then these could be used for a sampling scheme analogous to DN.

\paragraph{Breadth-first search (BFS):} The previous three methods randomly sampled nodes or edges. For the remaining methods, we will consider graph exploration approaches. The BFS algorithm begins by sampling a node at random. Then all neighbors of this sampled node are included in the sub-graph. Next, all neighbors of the neighbors are also included in the sub-sample. This process continues until we have sampled $qn$ nodes. BFS algorithms have long been employed in the computer science and operations research communities, and can be carried out in linear time in terms of the number of nodes and edges \citep{cormen2022introduction}. The algorithm has been used for various tasks such as computing shortest paths \citep[e.g.,][]{bundy1984breadth, beamer2013direction, burkhardt2021optimal}, graph partitioning \citep{jovanovic2017heuristic} and finding connected components of a graph \citep{jain2017adaptive}.

\paragraph{Depth-first search (DFS):} DFS can be considered as the complement of a BFS algorithm. This method also begins by randomly sampling a starting node. Then instead of including all neighbors of this starting node, it explores one (unexplored) edge to arrive at a new node. From here, a new unexplored edge is traversed, and this process continues until all of the current node's edges have been visited. At this point, the algorithm returns to the nearest node with unexplored edges and the process restarts. This procedure continues until $qn$ nodes have been sampled. DFS also runs linearly in terms of the number of nodes and edges \citep{cormen2022introduction} and has applications ranging from topological sorting of directed acyclic graphs and finding strongly connected components \citep{tarjan1972depth}, to multi-modal function optimization \citep{wang2022cooperative}. Finally, \cite{krishnamurthy2005reducing} studied the ability of BFS and DFS to preserve certain graph properties in the sub-samples, but found both performed poorly.

\paragraph{Random node-neighbor (RNN):} This approach shares similarities to BFS. As in BFS, a node is sampled at random and then all of its neighbors are also added to the sub-graph. Instead of looking at the neighbors of neighbors as in BFS, RNN samples a new node and includes all of the new node's neighbors. This continues until $qn$ nodes have been sampled. This approach is equivalent to sampling $1$-hop neighborhoods \citep[e.g.][]{mukherjee2021two}, and similar to an ego neighborhood except that the root node is also included in the sub-sample.
%If after the last step, $>qn$ nodes have been sampled, we randomly select $qn$ nodes from this set.

\paragraph{Random walk (RW):} In our final algorithm, we traverse the network through a random walk. Specifically, we begin by sampling a node uniformly at random. Then from this node, one of its neighbors is sampled at random, and this process continues until the walk has traversed $qn$ nodes. If the walk ends before reaching $qn$ unique nodes, then we restart the walk from a randomly selected node. Various works have studied the properties of RW-based sub-graphs \citep{ribeiro2012sampling, li2015random, chiericetti2016sampling}. In particular, \cite{lovasz1993random} shows that RW yields a uniform distribution on the edges, implying that RE and RW should perform similarly.\\

\noindent
Please see the Supplemental Materials for a figure which compares each of these algorithms on a small graph.

\section{Theoretical Results} \label{sec:theory}
In this section, we present theoretical results for each algorithm. As this paper's focus is on the sub-sampling step, we will primarily focus on the role of the sampling scheme on the theoretical results. Additionally, \cite{mukherjee2021two} devoted significant attention to the theoretical properties of $\mathsf{PACE}$, so we only briefly discuss these before developing new theoretical results for the CP divide-and-conquer algorithm.

\subsection{$\mathsf{PACE}$ algorithm}
The goal of the theoretical results for $\mathsf{PACE}$ is to show that the estimated clustering matrix $\hat \bC$, is ``close'' to the true clustering matrix $\bC$ which generated the network. Recall that $C_{ij}=1$, if nodes $i$ and $j$ are in the same community, and 0 otherwise. To formalize this notion, let $\tilde\delta(\bC,\hat \bC)$ be the difference between the true $\bC$ and estimate $\hat \bC$ where
$$
    \tilde\delta(\bC,\bC')
    =\frac{1}{n^2}||\bC-\bC'||_F^2
$$
and $||\cdot||_F$ is the Frobenius norm. Thus, $\tilde\delta(\bC,\hat \bC)$ is the mis-clustering rate of the $\mathsf{PACE}$ algorithm.  In the following theorem, the authors bound the expectation of this term under a general community detection algorithm and sampling scheme.\\

\noindent
{\bf Theorem 3.1 (Theorem 3.1 from \cite{mukherjee2021two}):} {\it Let graph $G$ have $K$ communities where community $k\in\{1,\dots,K\}$ contains $p_kn$ nodes, and define $p_{max}=\max\{p_1,\dots,p_K\}$ as the proportion of nodes in the largest community. Given some sampling scheme $\mathcal S$, let $S$ be a randomly chosen sub-graph of the network. Moroever, let $\hat \bC$ be the estimated clustering matrix returned by $\mathsf{PACE}$. Then the expected mis-clustering rate, $\mathbb E\tilde\delta(\bC,\hat \bC)$ can be bounded by:}
\begin{equation}\label{eq:pace_theory}
    \mathbb E\tilde\delta(\bC,\hat \bC)
    \leq \frac{B}{\beta n^2}\mathbb E||\hat \bC^{(S)}-\bC^{(S)}||^2_F + p_{max}\max_{i,j}\mathbb P(N_{ij}<\beta)
\end{equation}
{\it where $\hat \bC^{(S)}$ and $\bC^{(S)}$ are the estimated and true clustering matrices, respectively, restricted to sub-graph $S$.}\\

\noindent
There are two main parts to the bound in \eqref{eq:pace_theory}. The first term quantifies the performance of the community detection algorithm on a randomly chosen sub-graph $S$ which is mainly a function of the clustering algorithm. On the other hand, the second term measures how well the graph is covered by the sub-samples and depends only on the sub-sampling procedure. 

As our interest is on the sub-sampling step, we focus on this second term. Indeed, the authors of $\mathsf{PACE}$ provide two special cases of the results under the RN sampling scheme and ego neighborhood sampling. As we do not discuss ego neighborhoods, we omit discussion of this result. The following gives the results for RN sampling.\\

\noindent
{\bf Corollary 3.1 (Corollary 3.1 from \cite{mukherjee2021two}):} {\it Assume that RN is used to construct the sub-graphs. Then}
\begin{equation}
    p_{max}\max_{i,j}\mathbb P(N_{ij}<\beta) = O(e^{-\kappa Bp})
\end{equation}
{\it where $p=q^2(1+o(1))$, $\beta=\theta B p$ for $\theta\in(0,1)$ and $\kappa=(1-\theta)^2/2$.}\\

\noindent
This result shows that the mis-clustering error coming from the sub-sampling step decreases exponentially with $B$, the total number of sub-samples drawn, as well as $q$, the proportion of nodes in each sub-graph. As the authors note, deriving the specific result for other sub-sampling routines is difficult since the distribution of $N_{ij}$ is not simply $N_{ij}\sim \mathsf{Binomial}(B,q)$ as in the RN setting.

\subsection{CP algorithm}
Using similar ideas to that of \cite{mukherjee2021two}, we present novel theoretical results for the CP divide-and-conquer algorithm. As in the previous sub-section, we are interested in bounding the expected mis-classification error.

\subsubsection{General mis-classification result}

First, we derive the expression for the general mis-classification rate. Recall from Algorithm 2, our estimate $\hat{\bc}$ is defined as
$$
    \hat c_i = \frac1B\sum_{b=1}^B \hat c_i^{(b)}
$$
where $\hat{\bc}^{(b)}$ are the CP labels from sub-sample $b$, i.e., $\hat c_i^{(b)}=1$ if node $i$ was assigned to the core in sub-sample $b$, and 0 otherwise. Then we define the mis-clustering of the algorithm as
$$
    \delta(\hat\bc,\bc^*)
    = \frac1n ||\hat\bc-\bc^*||_2^2
    =\frac1n\sum_{i=1}^n (\hat c_i-c_i^*)^2
$$
where $\bc^*$ are the true CP labels, i.e., $c_i^*=1$ if node $i$ is a core node in the data-generating process, and 0 otherwise. The following result is analogous to Theorem 3.1 from \cite{mukherjee2021two}.\\

\noindent
{\bf Theorem 3.2:} {\it Given some graph $G$ with adjacency matrix $\bA$ and sampling scheme $\mathcal S$, let $S$ be a randomly chosen sub-graph. Let $\hat{\bc}$ be the estimated CP labels using the CP divide-and-conquer algorithm. Then the expected mis-clustering rate, $\mathbb E\delta(\hat\bc,\bc^*)$ can be bounded by}
\begin{equation}\label{eq:theo_cp}
    \mathbb E \delta(\hat\bc,\bc^*)
    \leq \frac1n\left(\mathbb E||\hat{\bc}^{(S)}-(\bc^*)^{(S)}||_2^2+\mathbb E\sum_{i=1}^n y_i^{(S)}\right)
\end{equation}
{\it where $\hat{\bc}^{(S)},(\bc^*)^{(S)}\in\{0,1\}^{qn}$ are the estimated and true CP labels, respectively, restricted to sub-graph $S$, and $y_i^{(S)}=1$ if node $i$ is in the core and was not sampled in sub-graph $S$, and 0 otherwise.}

Please see the Supplemental Materials for all proofs in this sub-section. Similar to the results of $\mathsf{PACE}$, our error bound has two terms; the first which primarily depends on the CP algorithm, and the second which only depends on the sub-sampling routine. Indeed, if the expected fraction of mis-classified nodes and the expected fraction of core nodes that are not sampled both go to 0 for large $n$, then this entire term will converge to 0. One noticeable difference between the two results is that Theorem 3.2 does not depend on $B$, the number of sub-samples, while Theorem 3.1 does. This discrepancy is because the CP divide-and-conquer algorithm divides the number of times a node was assigned to the core by $B$, rather than the number of times it was sub-sampled. If we had taken the latter approach, then $B$ would have appeared in the final result, along with a term counting the number of times a node was sampled.

We focus on the second term which is the expected number of core nodes which are not sampled in a given sub-graph. First, this term shows the importance of using a sub-sampling routine which samples core nodes with high probability. Indeed, if the probability of a core node being sampled increases, then this term decreases. Secondly, this term notably differs from the analogous term in the $\mathsf{PACE}$ theory. The CP algorithm divides the estimates by $B$, the total number of sub-samples, instead of the number of sub-graphs for which a node was sampled as in the $\mathsf{PACE}$ algorithm. Because of this, an error is guaranteed if a core node is not sampled. This choice will make the theory easier to derive, but puts us at risk of greater error. The key observation is that in real-world networks, the number of the core nodes is typically much smaller than the number of peripheral nodes. Indeed, in practice we found that dividing by $B$ instead of the number of times a node was sub-sampled led to improved performance. 

\subsubsection{Specific sub-sampling results}
The results of Theorem 3.2 are agnostic to the network-generating model, CP detection algorithm and sub-sampling scheme. By assuming a particular model, we can find the analytic expression for the second term in \eqref{eq:theo_cp} for the several of the sub-sampling schemes in Section \ref{sec:sub}.

We assume networks are generated with a CP structure from the stochastic block model (SBM) \citep{holland83}. For an adjacency matrix $\bA$, let $A_{ij}\stackrel{ind.}{\sim}\mathsf{Bernoulli}(\varrho_nP_{ij})$ where $P$ corresponds to a SBM that does not depend on $n$, and $\varrho_n=o(1)$ is a sparsity-inducing parameter. Furthermore, if $\bc^*$ corresponds to the true CP labels, then
$$
    P_{ij}
    =
    \begin{cases}
        p_{11} & \text{if }c_i^* = c_j^* = 1,\\
        p_{22} & \text{if }c_i^*=c_j^*=0,\\
        p_{12} & \text{otherwise}.\\
    \end{cases}
$$
If $p_{11}>p_{12}>p_{22}$, then this model generates networks with a CP structure, so we call it the CP-SBM. We may also assume that $\alpha_n=k/n$, the proportion of core nodes, is such that $\alpha_n=o(1)$ to model the empirical observation that the core is typically smaller than the periphery. In the following corollary, we find an analytic expression for the expected number of core nodes that are not sampled in each sub-graph, the second term in \eqref{eq:theo_cp}.\\

\noindent
{\bf Corollary 3.2:} {\it Let $\bA$ be generated from a CP-SBM with $n$ nodes and $k$ core nodes. Let $ S$ be a randomly chosen sub-graph of $\bA$ of size $q_nn$  where $q_n\in(0,1)$ may depend on $n$. Furthermore, let $y_i^{(S)}$ is the event that node $i$ is a core node and not sampled in $ S$. If RN is used to sample $ S$ and $q_n=O(1)$, then}
\begin{equation}\label{eq:cp_theory_rn}
        \mathbb E\frac1n\sum_{i=1}^n y_i^{(S)}
    = (1-q)\frac{k}{n}.
\end{equation}
{\it If $S$ is sampled via DN or RE and $q_n=o(1)$, then}
\begin{equation}\label{eq:cp_theory_dn}
    \mathbb E\frac1n\sum_{i=1}^n y_i^{(S)}
    \sim \left(1-q_n\frac{p_{12}}{p_{22}}\right)\alpha_n,
\end{equation}
{\it where $\sim$ denotes the asymptotic behavior for large $n$.}

The proof carefully computes the expected number of cores nodes that are sampled in each sub-graph. For RN, the number of core nodes sampled is exactly hyper-geometric distributed, meaning the result is non-asymptotic and does not require any assumptions on $q$. For DN and RE on the other hand, we are forced to rely on the node's expected degree, leading to the asymptotic result. The assumption $q_n=o(1)$ is needed to ensure that the probability of sampling the same node twice is negligible. It also ensures that there exists some $N$ such that for all $n\geq N$, $q_n\tfrac{p_{12}}{p_{22}}<1$ such that the expectation is always positive. Assuming $q_n=o(1)$ is reasonable as otherwise the number of nodes in the sub-sample would grow linearly with $n$. Given this assumption, the sub-sample size can still grow with $n$ but it must be sub-linear.

In both \eqref{eq:cp_theory_rn} and \eqref{eq:cp_theory_dn}, as size of the sub-graphs increases ($q_n$), this term decreases. Moreover, this expression depends on $\alpha_n=k/n$. Assuming that the periphery dominates the core, this entire term will go to zero regardless of the sampling scheme which theoretically validates our decision to divide the core estimates by $B$. Of course, the sub-sampling routine will still affect the performance for finite $n$.

It is also noteworthy that DN and RE yield identical expressions. This means these sampling schemes are equivalent in terms of the number of core nodes that are expected to be sampled in each sub-graph when the network is generated from a CP-SBM. This does not ensure that they will perform identically as the sub-sampling scheme affects the topology of the sub-graph $ S$ which may affect the first term in \eqref{eq:theo_cp}. 
Moreover, in a CP-SBM we have $p_{12}>p_{22}\implies p_{12}/p_{22}>1$, meaning that DN and RE have a uniformly smaller expected error than RN. Additionally, as the strength of the CP structure of the network increases, the error will decrease because $p_{12}/p_{22}$ increases. In other words, we expect a lower error rate if the network exhibits a stronger CP structure.

It is important to highlight that these results and discussion apply only to the second term in \eqref{eq:theo_cp}. It is possible that a sub-sampling method may not sample core nodes as often, but has a lower mis-classification rate on the sub-graph, leading to a smaller value of the first term in \eqref{eq:theo_cp}, or vice-versa. Finally, we briefly mention why we have not derived results for the other sub-sampler algorithms. For RNN, the difficulty is that the number of nodes sampled on each draw is different since we are sampling a node and its neighbors. BFS and DFS depend more heavily on the observed graph as opposed to the expected behavior of the nodes, complicating the analysis. Lastly, it is difficult to keep track of the visited nodes in RW. Deriving the results for these sub-sampling routines is another important avenue of future work.

\section{Simulation Study}\label{sec:sim}
In this section, we carry out extensive simulations of both divide-and-conquer algorithms. The goals of this section are to compare the performance of the algorithms under different sub-sampling schemes, as well as compare the divide-and-conquer algorithms against the original algorithm applied to the entire graph. Additionally, we hope to empirically validate the CP theoretical results. All experiments in this section were carried out in \texttt{R} (Version 4.4.2) on a Mac Mini with Apple M4 chip and 16 GB of memory parallelized using 9 cores. The code to replicate these simulations is available on the author's GitHub: \url{https://github.com/eyanchenko/subsamp}.

\subsection{Community detection simulations}
We begin by comparing the performance of $\mathsf{PACE}$ with different sub-sampling algorithms. By generating networks with a known community structure, we can compare the outputted membership vectors with the true clusters.

\subsubsection{Settings}
There are six simulations settings. In each, we generate networks, apply Algorithm 1 and compare the estimated membership vector with the true labels using the adjusted Rand index (ARI) \citep{rand1971objective, hubert1985comparing}.  The Rand index, $R(\hat\bc, \bc)$, provides a measure of concordance between the true, $\bc$, and estimated communities, $\hat\bc$. We define $c_i=k$ if node $i$ is in community $k$ for $k\in\{1,\dots,K\}$, and similarly, $\hat c_i=k$ if Algorithm 1 assigns node $i$ to community $k$. The Rand index is then computed as
$$
    R(\hat\bc, \hat\bc) = \frac{1}{{n\choose 2}}\sum_{i<j}^n \gamma_{ij}(\hat\bc,\bc)
$$
where
$$
    \gamma_{ij}(\hat\bc,\bc)
    =\begin{cases}
        1&\hat c_i=\hat c_j \text{ and } c_i=c_j,\\
        1&\hat c_i\neq \hat c_j \text{ and } c_i\neq c_j,\\
        0&\text{otherwise}.
    \end{cases}
$$
The adjustment in the ARI accounts for the fact that even a poor community detection algorithm could correctly classify some nodes just by chance. The maximum value of the ARI is 1 which indicates a perfect match between the estimated and true community labels (up to permutation of community labels).

We generate networks with a community structure by using a $K=2$-block SBM. Recall from Section \ref{sec:theory}, if $\bA$ is the corresponding adjacency matrix with data-generating model ${\bf P}$, then $A_{ij}\stackrel{ind.}{\sim}\mathsf{Bernoulli}(P_{ij})$ where $P_{ij}=p_{c_i, c_j}$. We take $p_{11}=p_{22}>p_{12}$ to generate networks with assortative community structure where $p_{11}=p_{22}$ and $p_{12}$ represent the intra-community and inter-community edge probabilities, respectively.  We set $p_{12}=0.01$ for all experiments in this sub-section. 

An overview of the simulation settings is available in Table \ref{tab:pace_settings}. Setting 1 looks at the effect of the intra-community edge probability where larger values indicated stronger community structure, and we vary $p_{11}=0.020$, $0.025$, $\dots$, $0.10$. Settings 2 and 3 increase the number of nodes in the network with $B$ held fixed and increasing with $n$, respectively, for $n=1\hspace{0.07cm}000$, $2\hspace{0.07cm}000$, $\dots$, $8\hspace{0.07cm}000$. The proportion of nodes, $\alpha$, in the first community is the subject of setting 4 where greater imbalance in the community sizes should make the task more difficult. We consider $\alpha=0.50$, $0.60$, $\dots$, $0.90$, $0.95$. Finally, settings 5 and 6 look at the hyper-parameters in the $\mathsf{PACE}$ model, namely the number of sub-graphs drawn and size of sub-graphs, respectively. In setting 7, $B$ varies from $B=100$, $250$, $500$, $1\hspace{0.07cm}000$, $2\hspace{0.07cm}500$, $5\hspace{0.07cm}000$, $10\hspace{0.07cm} 000$; while in setting 8, $qn$ varies from $qn=100$, $200$, $\dots$, $500$. For all parameter combinations, we generate 100 networks and record the average ARI and total computation time. In the Supplemental Materials, we also record the average computation time to detect the community structure on each sub-sample. Finally, we use the Fast Greedy algorithm \citep{Clauset:2004aa} as the base algorithm as in \cite{zhang2022distributed}. We also apply this algorithm to the entire network as a competing method to the divide-and-conquer approaches, labeled Full. In the Supplemental Materials, we repeat the same experiments using Walktrap \citep{Pons:2005aa} as the base algorithm, while also reporting the average sub-graph density for each algorithm across all settings.

\begin{table}[]
    \centering
    \begin{tabular}{c|ccccc}
        Setting & $n$ & $p_{11}$ & $\alpha$ & $B$ & $qn$ \\\hline
        1 & $5\hspace{0.07cm}000$ & ${\bf (0.02, 0.10)}$  & $0.75$ & $1\hspace{0.07cm}000$ & $250$\\
        2 & ${\bf(1\hspace{0.07cm}000, 8\hspace{0.07cm}000)}$ & $0.04$  & $0.75$ & $1\hspace{0.07cm}000$ & $250$\\
        3 & ${\bf(1\hspace{0.07cm}000, 8\hspace{0.07cm}000)}$ & $0.04$  & $0.75$ & ${\bf\frac12} \boldsymbol{n}$ & $250$\\
        4 & $5\hspace{0.07cm}000$ & $0.04$  & ${\bf(0.50,0.95)}$ & $1\hspace{0.07cm}000$ & $250$\\
        5 & $5\hspace{0.07cm}000$ & $0.04$  & $0.75$ & ${\bf(100, 10\hspace{0.07cm} 000)}$ & $250$\\
        6 & $5\hspace{0.07cm}000$ & $0.05$  & $0.75$ & $1\hspace{0.07cm}000$ & ${\bf (100, 500)}$\\
    \end{tabular}
    \caption{Simulation settings for $\mathsf{PACE}$ divide-and-conquer algorithm. $n$ is the number of nodes in the network and for settings 2 and 3 it varies from $n=1\hspace{0.07cm}000$, $2\hspace{0.07cm}000$, $\dots$, $8\hspace{0.07cm}000$; $p_{11}$ is the intra-community edge probability and for setting 1 it varies from $p_{11}=0.020$, $0.025$, $\dots$, $0.10$; $\alpha$ is the proportion of nodes in community 1 and in setting 4 varies from $\alpha=0.50$, $0.60$, $\dots$, $0.90$, $0.95$; $B$ is the number of sub-graphs sampled and in setting 7 varies from $B=100$, $250$, $500$, $1\hspace{0.07cm}000$, $2\hspace{0.07cm}500$, $5\hspace{0.07cm}000$, $10\hspace{0.07cm} 000$; and $qn$ is the number of nodes per sub-graph and in setting 8 varies from $qn=100$, $200$, $\dots$, $500$. The parameters which vary are highlighted in ${\bf bold}$.}
    \label{tab:pace_settings}
\end{table}

\subsubsection{Results}

\begin{figure}
    \centering
    \includegraphics[width=\linewidth]{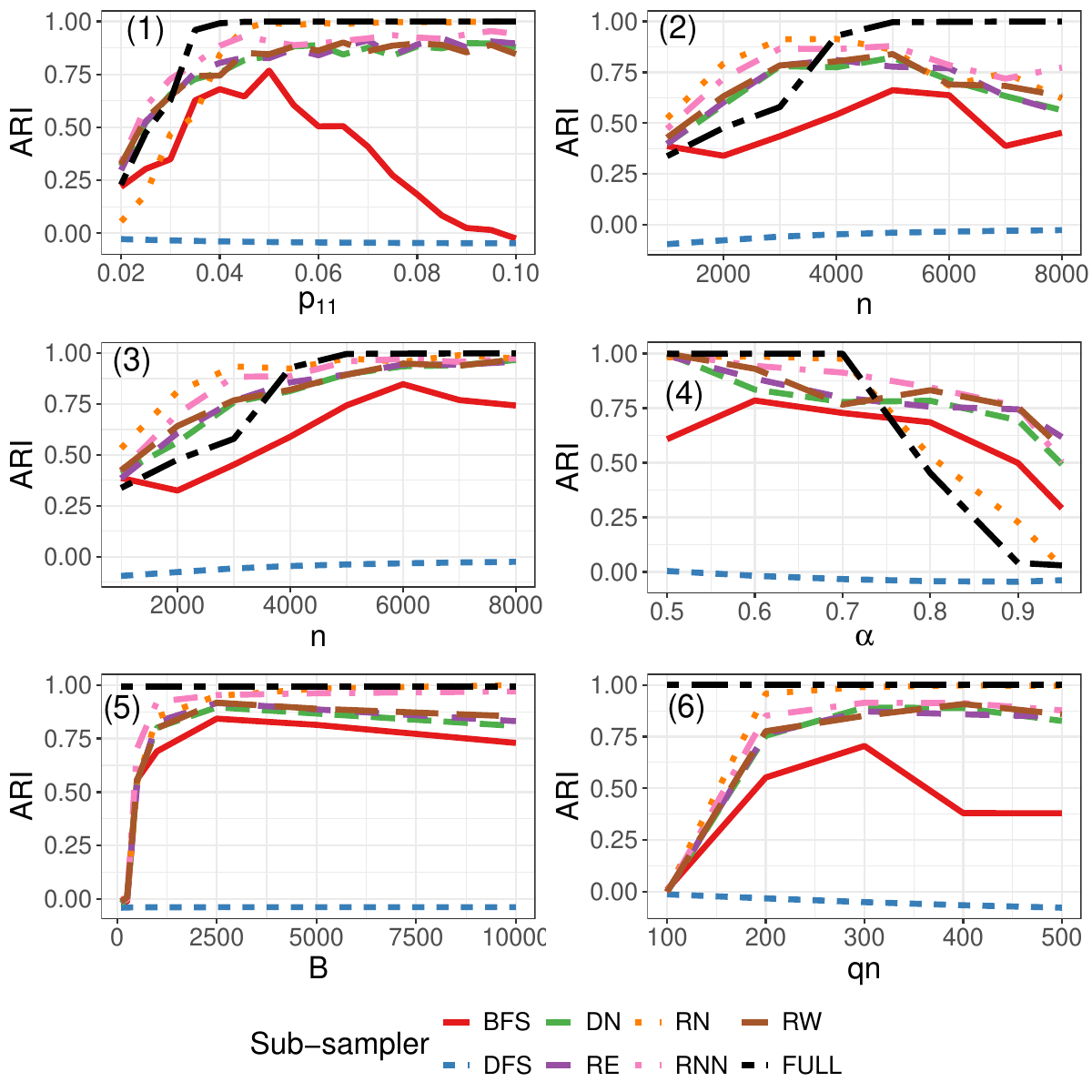}
    \caption{Community detection simulation results with Fast Greedy base algorithm. The number in the upper-left corner corresponds to the simulation setting (1-6).}
    \label{fig:com_res}
\end{figure}

\begin{figure}
    \centering
    \includegraphics[width=\linewidth]{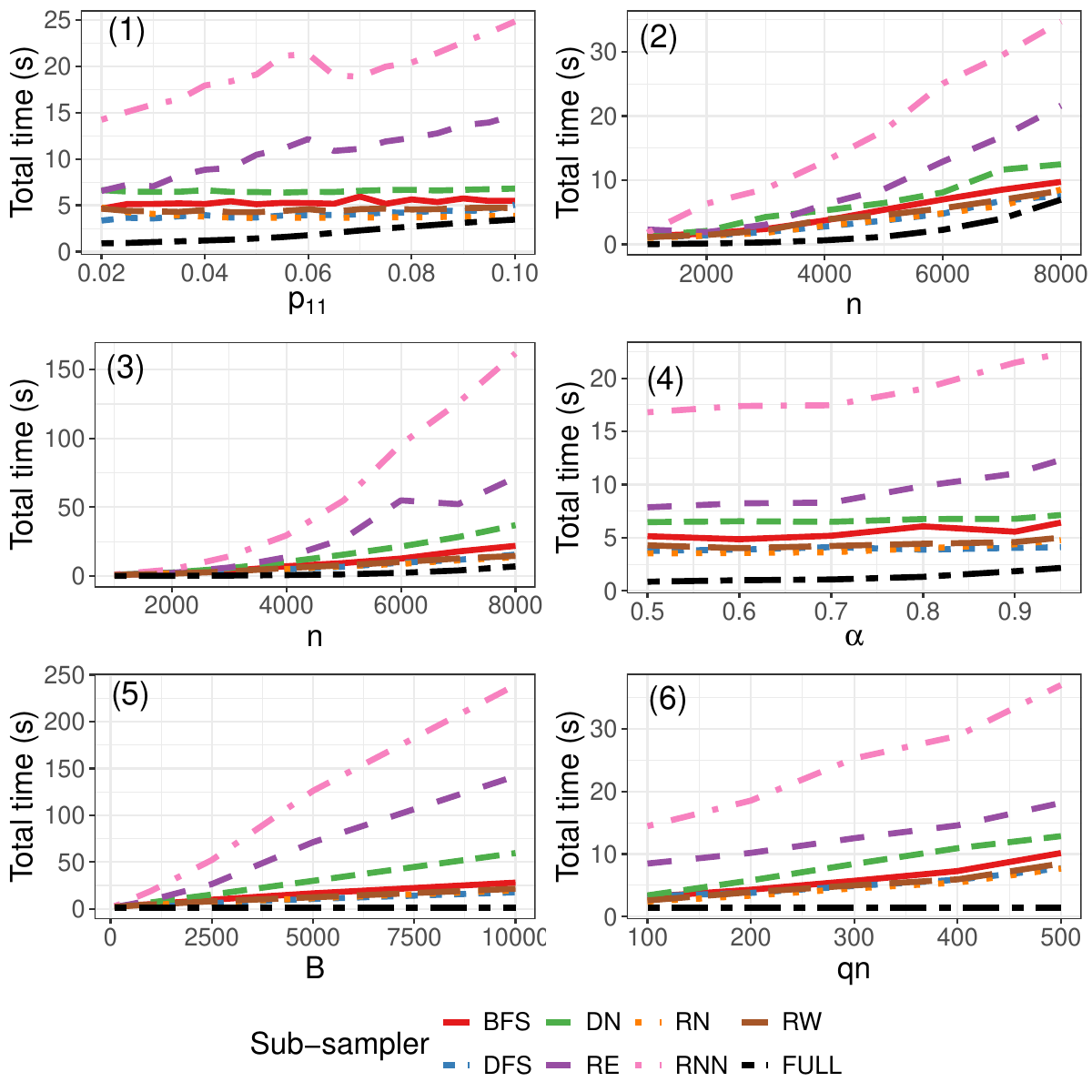}
    \caption{Community detection simulation run-time results with Fast Greedy base algorithm. The number in the upper-left corner corresponds to the simulation setting (1-6).}
    \label{fig:com_resT}
\end{figure}

The results are in Figures \ref{fig:com_res} and \ref{fig:com_resT}. In setting 1, save BFS and DFS, all methods yield an increasing ARI as $p_{11}$ increases. When $p_{11}\leq 0.04$, RNN has the largest ARI while RN does the best when $p_{11}>0.04$ among the divide-and-conquer methods. BFS improves at first before quickly dropping off, while DFS yields extremely low ARI values. Indeed, in all settings, it is clear that DFS performs poorly. Moreover, the algorithm applied to the full network yields the best performance when $p_{11}>0.03$. Turning to setting 2, the ARI increases with $n$ until around $n=4\hspace{0.07cm}000$ before decreasing for most methods. By $n=8\hspace{0.07cm}000$, these methods have experienced a drop off in performance. The full algorithm, on the other hand, has monotonically increasing performance with $n$, and yields the largest ARI values when $n\geq4\hspace{0.07cm}000$. When $B$ also increases with $n$ as in setting 3, DN, RE, RN, RNN and RW all have increasing ARI. RN and RNN perform slightly better and exceed or essentially match the ARI of Full for all $n$

The effects of the distribution of nodes between communities is evident in setting 4. Indeed, as the nodes are more disproportionately allocated between groups, each method decreases in performance with RN and Full being the most severely affected.  As $B$ increases in setting 5, all methods, save DFS, see an improvement in performance. Interestingly, while RN and RNN plateau at an ARI of 1, DN, RE, RW and BFS all begin to yield slightly decreasing values of ARI for $B>2\hspace{0.07cm}500$. Finally, setting 6 shows that as the number of nodes in the sub-samples increases, the communities are easier to detect for all methods except BFS and DFS. Among the divide-and-conquer methods, for most values of $q$, RN has the largest ARI followed by RNN. Note that the results for Full are constant in the last two settings since $B$ and $q$ do not appear as parameters in the full algorithm.

The computation time for each approach is in Figure \ref{fig:com_resT}. Somewhat surprisingly, Full yields the fastest computation time across all settings. The different $\mathsf{PACE}$ algorithms have similar computation times, with the exception of RNN, which consistently yields the slowest times. This is likely because this sub-sampling algorithm was written by hand, while the other sub-samplers were implemented from the $\texttt{igraph}$ package \citep{igraph}.

\subsubsection{Discussion}

The first major takeaway from the simulation study is the strong performance of the base algorithm applied to the entire network. Not only did it yield the best detection results across many settings (as we might have expected), but it was also the fastest algorithm as well. The advantage of the divide-and-conquer approach should be that it improves computation time. If the base algorithm is fast enough, as in this case, then evidently it does not. In \cite{mukherjee2021two}, the authors primarily discuss using spectral clustering or likelihood-based method as the base algorithm, which seem to be slower than Fast Greedy. In this case, $\mathsf{PACE}$ will yield computational advantages. For example, in the Supplemental Materials where we use Walktrap as the base algorithm, the divide-and-conquer approaches are indeed faster. This implies that depending on the choice of base algorithm, the advantages of $\mathsf{PACE}$ may differ. Even though $\mathsf{PACE}$ is slower, there are several settings where it yields larger ARI values, in particular for small $p_{11}$ (setting 1), small $n$ (settings 2 and 3) and highly disproportionate communities (setting 4). The similarity between these settings is that the community structure is fairly weak, either because $p_{11}$ and $p_{12}$ are close, the network is small, or one community dominates the network. In such cases, it could be that the sub-sampling approach is averaging out some of the noise in the network, yielding better identification results.

Looking only at the divide-and-conquer approaches, it appears that RN and RNN perform the best with RW, RE and DN also yielding good results. Even more clearly, the results demonstrate that BFS and DFS are poor samplers to use with the $\mathsf{PACE}$ algorithm. This is likely because the sub-graphs resulting from them are dissimilar from the original graph and do not preserve the community structure. This result is in agreement with \cite{krishnamurthy2005reducing} which showed poor performance by both exploration samplers. Setting 2 reveals the importance of increasing the number of sub-samples as $n$ increases, even though the theoretical results depend on a fixed $B$. Settings 5 and 6 show that $B$ and $q$ must be carefully chosen since small values can severely hinder performance. Finally, all methods suffer when the nodes are unequally distributed between communities. This is not surprising as the identification problem is much more challenging in this setting. Indeed, the presence of small communities likely requires increasing in the number of nodes per sub-sample in order to capture this structure. In the Supplemental Materials, we show that conclusions are similar when using Walktrap as the base community detection algorithm.

\subsection{Core-periphery detection simulations}
Similar to the previous sub-section, we now compare the performance of the various sub-sampling routines from Section \ref{sec:sub} on the CP divide-and-conquer algorithm.

\subsubsection{Settings}
We again look at six simulation settings. For each setting, we generate a network, apply Algorithm 2 and compute the area under the receiver operator curve (AUC) between the outputted proportions and true labels. CP identification is fundamentally about classifying nodes as core or periphery and AUC is a popular metric for such tasks \citep[e.g.,][]{Kojaku2018, naik2021}. The base CP detection algorithm is a greedy algorithm based on Algorithm 1 from \cite{yanchenko2022divide}. Similar to the community detection simulations, we also apply this algorithm to the entire network as a benchmark for comparison (Full). The networks are generated the same way as in the community detection simulations except now $p_{11}>p_{12}>p_{22}$ where $p_{11},p_{12},p_{22}$ represent the core-core, core-periphery, and periphery-periphery edge probabilities, respectively. For each setting, we fix $p_{22}=0.001$ and set $p_{12}=\frac12p_{11}$.

\begin{table}[]
    \centering
    \begin{tabular}{c|ccccc}
        Setting & $n$ & $p_{11}$ & $\alpha$ & $B$ & $qn$ \\\hline
        7 & $5\hspace{0.07cm}000$ & ${\bf (0.002, 0.020)}$  & $0.01$ & $1\hspace{0.07cm}000$ & $100$\\
        8 & ${\bf(1\hspace{0.07cm}000, 8\hspace{0.07cm}000)}$ & $0.004$  & $0.01$ & $1\hspace{0.07cm}000$ & $100$\\
        9 & ${\bf(1\hspace{0.07cm}000, 8\hspace{0.07cm}000)}$ & $0.004$  & $0.01$ & ${\bf\frac12} \boldsymbol{n}$ & $100$\\
        10 & $5\hspace{0.07cm}000$ & $0.004$  & ${\bf(0.002,0.30)}$ & $1\hspace{0.07cm}000$ & $100$\\
        11 & $5\hspace{0.07cm}000$ & $0.004$  & $0.01$ & ${\bf(100, 10\hspace{0.07cm} 000)}$ & $100$\\
        12 & $5\hspace{0.07cm}000$ & $0.004$  & $0.01$ & $1\hspace{0.07cm}000$ & ${\bf (50, 500)}$\\
    \end{tabular}
    \caption{Simulation settings for CP divide-and-conquer algorithm. $n$ is the number of nodes in the network and for settings 2 and 3 it varies from $n=1\hspace{0.07cm}000$, $2\hspace{0.07cm}000$, $\dots$, $8\hspace{0.07cm}000$; $p_{11}$ is the core-core edge probability and for setting 1 it varies from $p_{11}=0.002$, $0.003$, $\dots$, $0.020$; $\alpha$ is the proportion of nodes in the core and in setting 4 varies from $\alpha=0.002$, $0.004$, $\dots$, $ 0.008$, $0.01$, $0.02$, $\dots$, $0.05$, $0.10$, $0.20$, $0.30$; $B$ is the number of sub-graphs sampled and in setting 7 varies from $B=100$, $250$, $500$, $1\hspace{0.07cm}000$, $2\hspace{0.07cm}500$, $5\hspace{0.07cm}000$, $10\hspace{0.07cm} 000$; and $qn$ is the number of nodes per sub-graph and in setting 8 varies from $qn=50$, $100$, $250$, $500$. The parameters which vary are highlighted in ${\bf bold}$.}
    \label{tab:cp_settings}
\end{table}

The simulation settings are summarized in Table \ref{tab:cp_settings}.   In setting 7, we vary $p_{11}=0.002$, $0.003$, $\dots$, $0.020$ where larger values correspond to a stronger CP structure. Settings 8 and 9 look at the effect of increasing the network size where the former fixes $B$ while the latter allows $B$ to increase with $n$, i.e., $B=\frac12n$. Both settings consider $n=1\hspace{0.07cm}000$, $2\hspace{0.07cm}000$, $\dots$, $8\hspace{0.07cm}000$. The effect of the core size, $\alpha$, is studied in setting 10 for $\alpha=0.002$, $0.004$, $\dots$, $ 0.008$, $0.01$, $0.02$, $\dots$, $0.05$, $0.10$, $0.20$, $0.30$. The choice of hyper-parameters is investigated in settings 11 and 12, varying the values of $B$ and $q$, respectively. We vary $B=100$, $250$, $500$, $1\hspace{0.07cm}000$, $2\hspace{0.07cm}500$, $5\hspace{0.07cm}000$, $10\hspace{0.07cm} 000$ and $qn=50$, $100$, $250$, $500$ in these settings, respectively.  Each setting is repeated for 100 Monte Carlo replications and the average AUC and total run time are reported. In the Supplemental Materials, we also report the average detection time per sub-sample.

\subsubsection{Results}
\begin{figure}
    \centering
    \includegraphics[width=\linewidth]{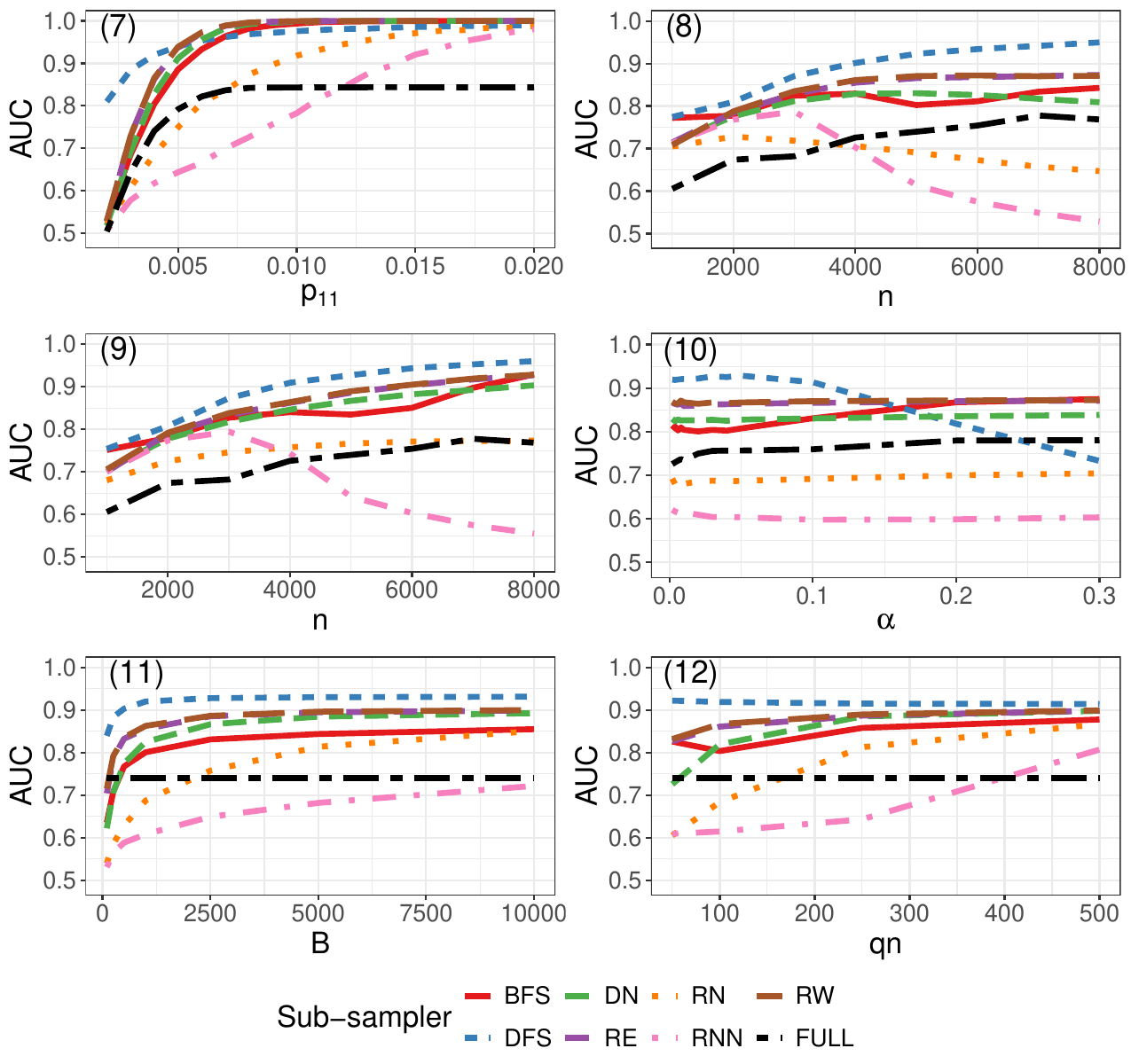}
    \caption{Core-periphery simulation accuracy results. The number in the upper-left corner corresponds to the simulation setting (7-12).}
    \label{fig:cp_res}
\end{figure}

The results are in Figure \ref{fig:cp_res}. In setting 7, as $p_{11}$ increases, all sub-sampling schemes have a non-decreasing AUC. BFS, DN, RE and RW all perform nearly identically with DFS performing the same once $p_{11}\geq 0.005$. On the other hand, RN and RNN yield worse performance, while Full's AUC plateaus before reaching 1. Next, it is helpful to compare settings 8 and 9 together since they both look at increasing $n$. Similar to setting 7, DFS yields the best results with BFS, DN, RE and RW yielding similar performance. RN and RNN consistently yield the lowest AUC values in both settings. In addition, we notice that if $B$ is fixed and $n$ increases (setting 8), the AUC values start to decrease for larger $n$, most notably for RN and RNN. In setting 9, however, where $B$ increases with $n$, all methods save RNN have a generally monotonically increasing AUC with increasing $n$. In both of these settings, the divide-and-conquer algorithms tend to outperform Full.

Moving to setting 10, the performance of DFS noticeably decreases as the core size increases. All other methods have roughly constant AUC values. In setting 11, we see that increasing the number of sub-sampled nodes beyond one or two thousand yields minimal returns for DFS, BFS, DN, RE and RW. RN, however, has monotonically improving AUC with increasing $B$ while RNN also improves with larger $B$. These trends are similar in setting 12. Increasing the number of nodes in the sub-samples only seems to noticeably improve RN and RNN; the other methods remain relatively unchanged, especially once $qn>50$. 

\begin{figure}
    \centering
    \includegraphics[width=\linewidth]{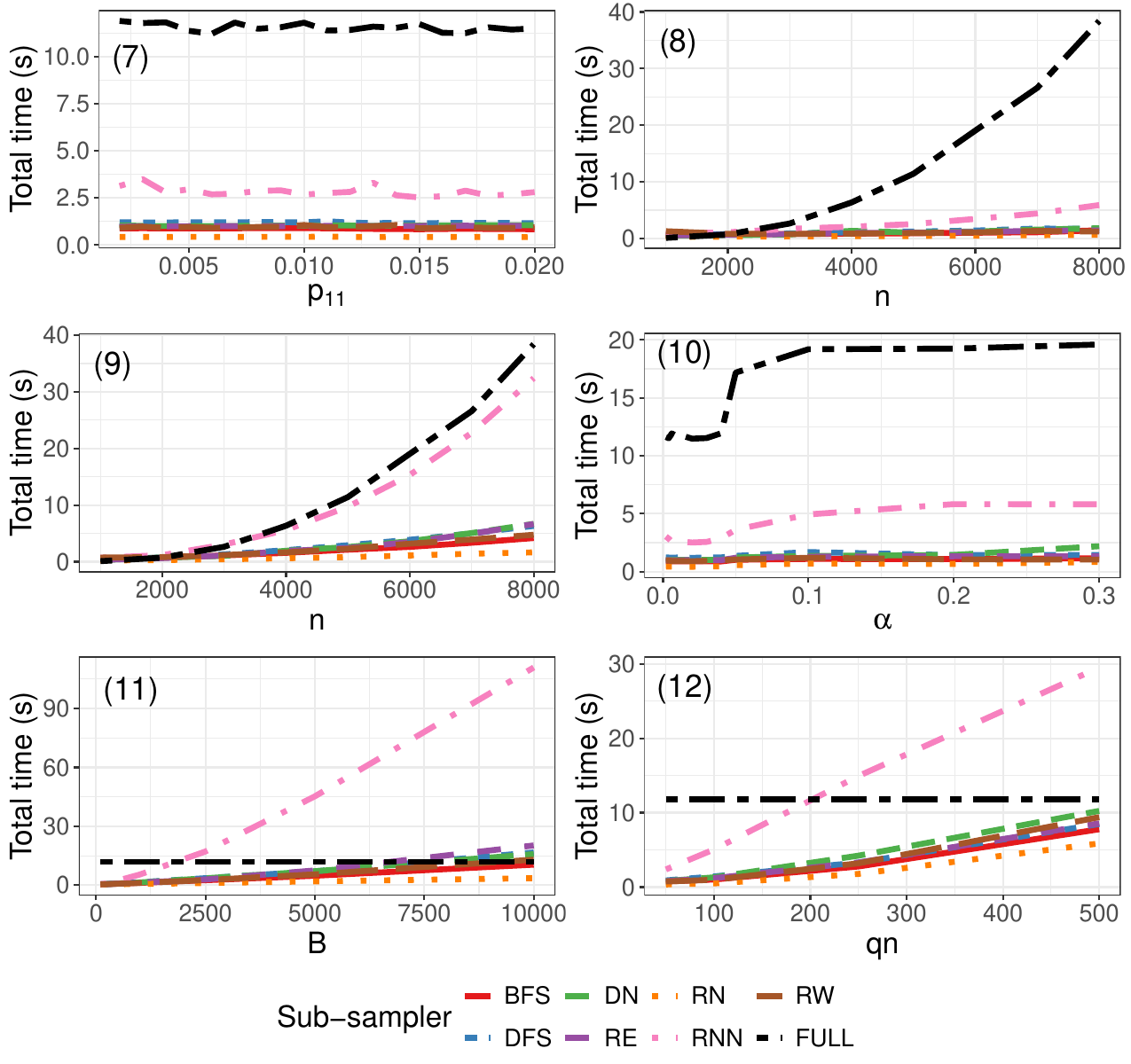}
    \caption{Core-periphery simulation run-time results. The number in the upper-left corner corresponds to the simulation setting (7-12).}
    \label{fig:cp_time}
\end{figure}

We also report the computation time for each approach in Figure \ref{fig:cp_time}. In settings 7, 8, 9, and 10, all methods are significantly faster than the full algorithm (save RNN in 9). From setting 11, if the number of sub-samples is large, then the divide-and-conquer algorithm's computation time is comparable to that of the full algorithm. RNN again consistently yields the slowest times for the divide-and-conquer approaches.

\subsubsection{Discussion}
There are several takeaways from these simulations. Surprisingly, the divide-and-conquer algorithms outperform the original algorithm applied to the full network, both in terms of AUC and computing time. We expected there to be computational benefits to the sub-sampling scheme, but the superior AUC is somewhat unexpected. We postulate that this is because the sub-sampling step is able to de-noise the results. By averaging the CP results over many sub-graphs, it seems that any outlying behavior of nodes on the original graph can be removed and/or a weak CP structure can be detected. Next, it is clear that BFS, DN, RE and RW are the best sub-sampling methods for CP identification. DFS also performs well but its behavior for increasing core size is somewhat concerning. On the other hand, RN and RNN display the worst performance. Since the mis-classification rate is lower when core nodes are sampled more frequently, the best methods will have a higher probability of sampling core nodes. Indeed, both RN and RNN sample core nodes less frequently than BFS, DN, RE and RW. This also aligns with the theoretical results which showed that RE and DN should perform similarly, and that both of these methods should outperform RN. Additionally, RW and RE results are almost indistinguishable across all settings, as expected based on \cite{lovasz1993random}. Continuing on the theoretical thread, we note that the theory in Section \ref{sec:theory} assumes that the number of sub-graphs, $B$, is fixed. In setting 8, however, we saw that this assumption can lead to poor performance empirically. Indeed, settings 8 and 9 seem to indicate that $B$ must be increased in larger networks, as there are more nodes to sample. Lastly, these results have implications on the choice of hyper-parameters. In settings 11 and 12, we found that increasing $B$ and $q$ only leads to marginal gains in performance, particularly for BFS, DN, RE and RW. This means that these values can be set relatively low, increasing the computational speed. RN and RNN require larger $B$ and $q$ to match the other method's performance, further demonstrating their weakness for this task.

%Whatever sampling method performs the best, we need to have an understanding of why it is doing best. One idea is to look at what sort of sub-graphs it is forming and see how closely they approximate the original graph. Could look at them individually. But could also sort the sub-sampled graphs in terms of node's degree and then plot the ``average'' sub-sampled graph. We can see how this compares to the original graph visually. Could also compute some network metrics on the sub-graphs such as $\rho$ (CP parameter), clustering coefficient, etc. and compare with original graph. Ideally, the sub-sampling schemes with best performance will yield similar metrics to the original graph.

\section{Real-data analysis}\label{sec:real}
In this section, we consider real-world networks from various domains to study the performance of the algorithms under different settings and types of network. The goal is to understand how the choice of sub-sampler affects the results of the divide-and-conquer algorithms. Moreover, we are interested in considering the theoretical results on non-synthetic data.

\subsection{Community detection}
We apply the $\mathsf{PACE}$ algorithm to two different real-world networks. This first is {\it Political Blogs} \citep{adamic::2005aa} ($n=1\hspace{0.07cm}490, m=16\hspace{0.07cm}715$) where nodes represent blogs and edges arise if one blog references another. The blogs are divided between liberal and conservative, so we set $K=2$, and keep only the largest connected component, yielding $n=1222$. This network was chosen since there are ground-truth labels (conservative or liberal) that we can compare with the estimated labels, and also because it was considered in the original $\mathsf{PACE}$ paper. The second is {\it Facebook} \citep{leskovec2012learning} ($n=4\hspace{0.07cm}039,m=88\hspace{0.07cm} 234$) where nodes are users connected with an edge if they are friends. This is an ego network with ten egos, so we set $K=10$. An ego refers to the sub-graph of all friends of a single user. This network was chosen since the number of communities is known, a requirement for $\mathsf{PACE}$. We remove all edge weights and directions as well as self-loops for both networks.

We apply the $\mathsf{PACE}$ algorithm to each network using Fast Greedy \citep{Clauset:2004aa} as the base community detection algorithm. We report the proportion of nodes in the largest community, $\hat\alpha$, as well as the modularity of the returned node membership vector, $Q$ \citep{newman2006modularity}. Modularity is a popular measure of the strength of community structure where a large value indicates a better community assignment \citep[e.g.,][]{newman2004finding, bickel2009nonparametric}. For a network $\bA$ and community labels $\bc$, it is computed as
$$
    Q({\bf A}, \bc)
    =\frac1m\sum_{i<j}\left(A_{ij}-\frac{d_id_j}{2m}\right)\mathbb I(c_i=c_j)
$$
where $d_i=\sum_jA_{ij}$ is the degree of node $i$. Since we have the ground truth labels (which we call the Oracle) for the {\it Political Blogs} network, we can also compute the ARI between the true and estimated labels. Finally, we set $q=250/n$ for each network, and $B=1\hspace{0.07cm}000$ and $5\hspace{0.07cm}000$ for {\it Political Blogs} and {\it Facebook}, respectively. 

The results are in Table \ref{tab:pace_real}. For {\it Political Blogs}, RN yields both the largest modularity as well as most similar labels to the ground-truth. RNN does the next best, with RE also performing similarly. On the other hand, DFS yields substantially lower modularity and ARI values, likely due to its much greater imbalance in community labels (large $\hat\alpha$). The results are similar for {\it Facebook}, where RN has the largest modularity and RNN provides the second-largest. In this example, both BFS and DFS provided imbalanced class labels, leading to lower modularity values.

\begin{table}[]
    \centering
    \begin{tabular}{l|ccc|cc}
    \multicolumn{1}{c|}{} &
    \multicolumn{3}{|c|}{Political Blogs} & \multicolumn{2}{|c}{Facebook}\\
    Method & $\hat\alpha$  & $Q$ & ARI & $\hat\alpha$ & $Q$ \\\hline
    Oracle & 0.52 & 0.41 &--  & -- & -- \\
    Full & 0.52 & 0.43 & 0.78 & 0.24 & 0.78\\ \hline
    RN & 0.53 & 0.42 & 0.81 & 0.46 & 0.74\\
    DN & 0.63 & 0.40 & 0.43 & 0.57 & 0.63\\
    RE & 0.66 & 0.41 & 0.50 & 0.64 & 0.55 \\
    BFS & 0.62 & 0.39 & 0.46 & 0.86 & 0.04\\
    DFS & 0.90 & 0.08 & 0.06 & 0.99 & 0.00\\
    RNN & 0.63 & 0.42 & 0.58 & 0.66 & 0.69\\
    RW & 0.62 & 0.41 & 0.45 & 0.72 & 0.55
    \end{tabular}
    \caption{Real data analysis results for PACE community structure divide and conquer algorithm. $\hat\alpha$ is the proportion of nodes in the largest community, $Q$ is the modularity and ARI is the adjusted Rand index between the estimated and ground-truth community labels.}
    \label{tab:pace_real}
\end{table}

\subsection{Core-periphery detection}
We now compare the sub-sampling routines using the CP divide-and-conquer algorithm on real-world networks: {\it Airport} \citep{igraph} is a network where nodes correspond to airports and edges to flights between airports ($n=755$, $m=4\hspace{0.07cm}623$); and {\it Twitch} \citep{rozemberczki2019, snapnets} has nodes which are users and edges represent friendships ($n=168\hspace{0.07cm} 114$, $m=6\hspace{0.07cm}797\hspace{0.07cm} 557$). Edge weights and directions were again removed, along with self-loops.

For each network, we report the size of the core, $k$, and corresponding Borgatti and Everett (BE) metric value for the optimal core returned by the algorithm, computed using \eqref{eq:BE}. The BE values quantifies the strength of core-periphery structure detected by the algorithm, with larger values implying a better core was found. In the simulation settings, we simply used the core proportions as a measure of a node's coreness, but for the real data, we desire binary CP labels. To convert the core proportions to binary labels, we sort the nodes in descending order of their proportions. Then one by one, starting with the node with the largest proportion, we add this node to the core and compute the BE metric. We continue this process and select the core corresponding to the largest BE metric.

While the airport network is relatively small, this allows us to find the CP labels using the base algorithm without the divide-and-conquer step. Recall that the base algorithm is based on Algorithm 1 from \cite{yanchenko2022divide}. We label this as ``Full'' in our results and provides us with a useful benchmark. Finally, we note the following values of $q$ and $B$ that were used for each network: {\it Airport}, $q=100/n$ and $B=5\hspace{0.07cm}000$; and {\it Twitch}, $q=500/n$ and $B=25\hspace{0.07cm} 000$.

\begin{figure}
    \centering
    \includegraphics[width=\linewidth]{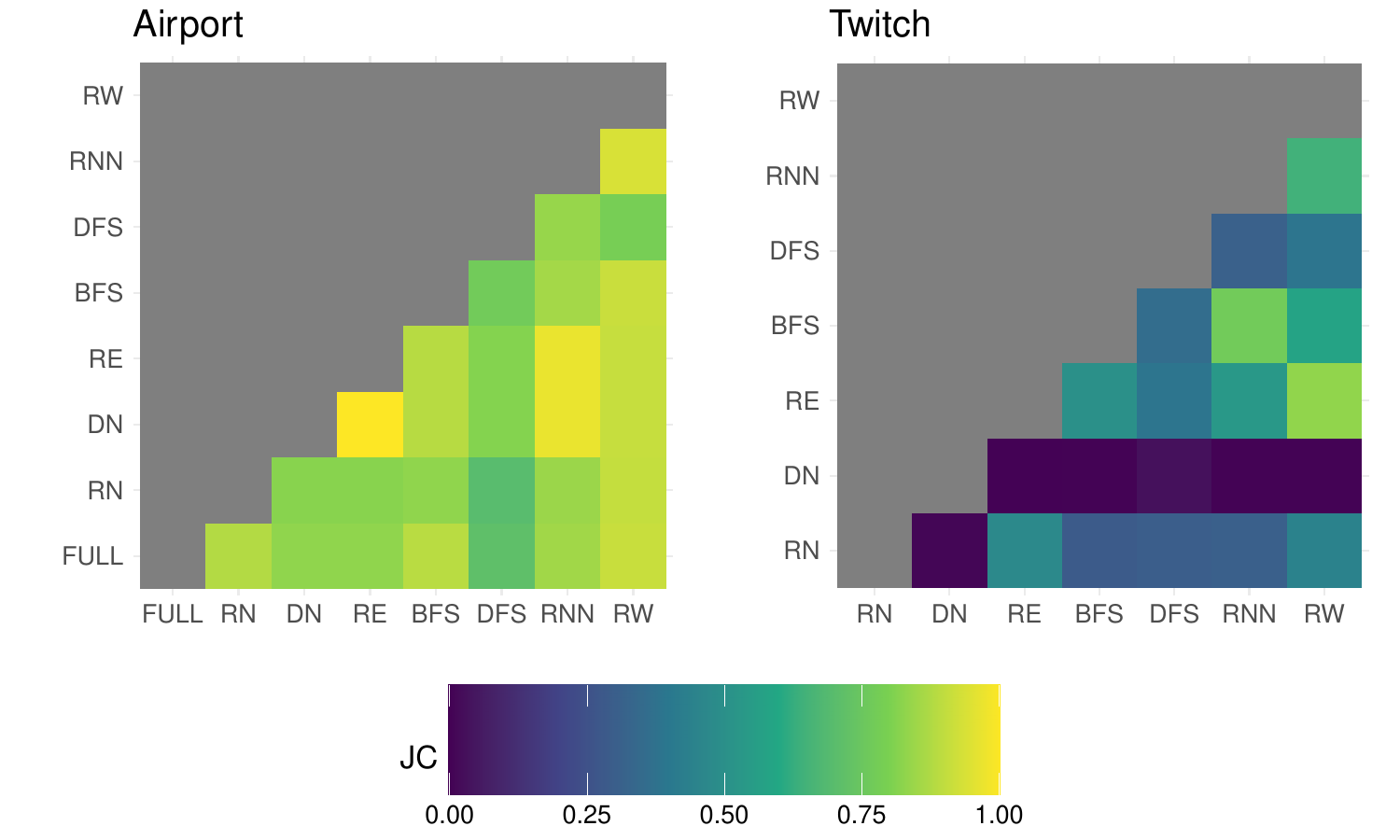}
    \caption{Comparison of cores returned using different sub-sampling algorithms. The color of the square corresponds to the Jacaard coefficient between the two core sets with lighter color meaning more similarity.}
    \label{fig:cp_real}
\end{figure}

In Table \ref{tab:cp_real}, we report the core size and BE metric for the detected cores, and in Figure \ref{fig:cp_real}, we compare the cores returned by the different algorithms. Specifically, for each sub-sampling pair, we compute the Jacaard coefficient (JC) \citep{liben2003link} between the returned core sets. If $S_1$ and $S_2$ are two sets, then the JC is computed as
$$
    JC(S_1,S_2)
    =\frac{|S_1\cap S_2|}{|S_1\cup S_2|}.
$$
In words, the JC is the proportion of total elements in the two sets which are shared. It takes values from 0 to 1 with 1 indicating that $S_1$ and $S_2$ are identical. In this way, the more similar the cores are, the larger the JC will be. 

For {\it Airport}, all method return similar core sizes, cores and BE metric values as that of Full. Moreover, the cores for DN and RE are exactly the same (JC of 1), supporting the results from Corollary 3.2. For {\it Twitch}, RE and RW yield the largest BE values, with RN, BFS and RNN also having similar values. DN yields a vastly different core size and BE value, while DFS does not perform as well. For this network, RW and RE have the most similar cores, which supports \cite{lovasz1993random}. On the other hand, DN and RE's results are quite dissimilar, implying that the SBM assumption of Corollary 3.2 may be necessary for the results to hold.

\begin{table}[]
    \centering
    \begin{tabular}{l|cc|cc}
    \multicolumn{1}{c|}{} &
    \multicolumn{2}{|c|}{Airport} & \multicolumn{2}{|c}{Twitch}\\
    Method & $k$ & BE & $k$ & BE  \\\hline
    Full & 33 & 0.24  & -- & --\\\hline
    RN & 29 & 0.24 & 275 & 0.08 \\
    DN & 31 & 0.23 & 164 & 0.00\\
    RE & 31 & 0.23  & 170 & 0.08\\
    BFS & 35 & 0.23  & 88 & 0.08\\
    DFS & 27 & 0.23 & 155 & 0.08 \\
    RNN & 30 & 0.23  & 91 & 0.08\\
    RW & 32 & 0.24 & 141 & 0.08
    \end{tabular}
    \caption{Real data analysis results for core-periphery divide and conquer algorithm. $k$ refers to the number of nodes assigned to the core by the respective algorithms, and BE is the value of the Borgatti and Everett metric \eqref{eq:BE} corresponding to the optimal labels.}
    \label{tab:cp_real}
\end{table}

\section{Conclusion}\label{sec:conc}
Graph sub-sampling has long been recognized as an important task by network scientists and computer scientists, but it has yet to draw as much attention from statisticians. In this work, we offer an empirical and theoretical comparison of seven graph sub-sampling algorithms from a statistical angle by applying these methods to divide-and-conquer algorithms for two important meso-scale network features: community structure and core-periphery structure. 

Based on our findings, we recommend random node sampling when trying to detect community structure. This method had some of the best performance in the simulated networks as well as on the real-world networks. Moreover, random node sampling was studied theoretically in \cite{mukherjee2021two} and is the easiest sub-sampling method to implement. The reason for this good performance could be because random node sampling can easily sample across the network without getting ``stuck'' in any one part. Of course, random node sampling, as well as the other methods, will suffer if there are communities with only a few nodes. In this case, it is important to increase the size of the sub-graphs and/or the number of iterations. Randomly sampling nodes proportional to their degree as well as random node neighbor sampling also provided good results on the synthetic and real-world networks. Breadth-first and depth-first search algorithms, however, should be avoided for this task. Additionally our results showed that in many cases, the divide-and-conquer algorithm is not needed at all, as the base algorithm can yield good detection results in a short amount of time.

As for core-periphery identification, there was no clear winner, but samplers which selected core nodes with a higher probability consistently performed the best. Random edge sampling and random walk both yielded good results across the simulated and real-world data.  On the other hand, random node sampling and random node neighbor performed quite poorly in these settings, aligning with our theoretical results. The real-data results show that many different cores can yield similar objective function values, implying the objective function surface is relatively ``flat.'' For CP identification, our results imply that the divide-and-conquer algorithm should always be preferred to the base algorithm as it is faster and is better at identifying the core.

It is important to note that these results were based on our simulated and real-world data experiments and may not generalize to other settings. That being said, the contrasting performance of sub-sampling routines on the different identification tasks is revealing. Random node and random node neighbor sampling performed the best for the community detection problem, but were the two weakest methods for core-periphery identification. On the other hand, breadth-first and depth-first searches are good approaches for identifying core-periphery structure, but not for community structure. These results highlight the importance of tailoring the choice of sub-sampling algorithm to the particular problem. Indeed, our CP theory revealed the optimal feature for a sampler is a high probability of sampling core nodes. Deriving similar theory for future problems will help researchers choose the best methods, or even may even drive them to derive novel sub-sampling schemes for the specific problem.

\section*{Acknowledgments}
The author would like to thank the Associate Editor and two anonymous referees whose comments and suggestions greatly improved the manuscript.

\section*{Statements and Declarations}
The author declares no competing interests nor have any funding to report.

\bibliographystyle{apalike}
\bibliography{refs}

\clearpage

\begin{center}
    {\Large Supplemental Materials}
\end{center}

\section*{Sub-sampling algorithms}
\begin{figure}
    \centering
    \begin{tabular}{cc} % 4 rows, 2 columns
        \begin{subfigure}{0.45\textwidth}
            \centering
            \includegraphics[width=\linewidth]{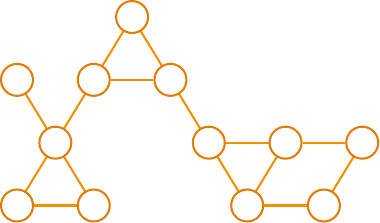}
            \caption{Original graph}
        \end{subfigure} &
        \begin{subfigure}{0.45\textwidth}
            \centering
            \includegraphics[width=\linewidth]{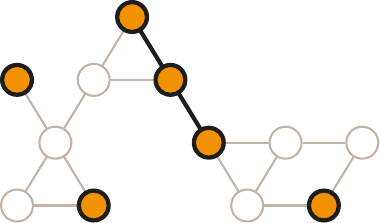}
            \caption{Random node}
        \end{subfigure} \\
        
        \begin{subfigure}{0.45\textwidth}
            \centering
            \includegraphics[width=\linewidth]{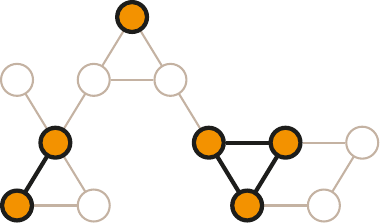}
            \caption{Degree node}
        \end{subfigure} &
        \begin{subfigure}{0.45\textwidth}
            \centering
            \includegraphics[width=\linewidth]{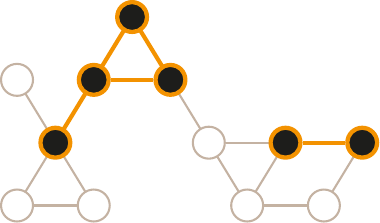}
            \caption{Random edge}
        \end{subfigure} \\
        
        \begin{subfigure}{0.45\textwidth}
            \centering
            \includegraphics[width=\linewidth]{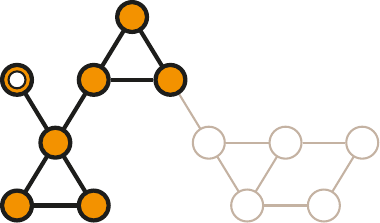}
            \caption{Breadth-first search}
        \end{subfigure} &
        \begin{subfigure}{0.45\textwidth}
            \centering
            \includegraphics[width=\linewidth]{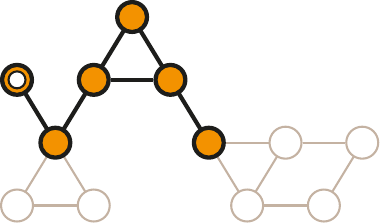}
            \caption{Depth-first search}
        \end{subfigure} \\
        
        \begin{subfigure}{0.45\textwidth}
            \centering
            \includegraphics[width=\linewidth]{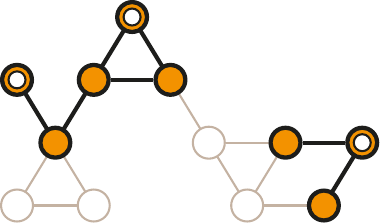}
            \caption{Random node neighbor}
        \end{subfigure} &
        \begin{subfigure}{0.45\textwidth}
            \centering
            \includegraphics[width=\linewidth]{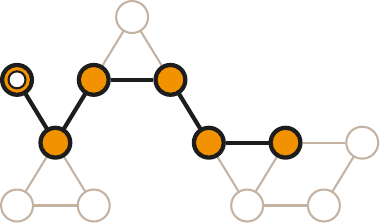}
            \caption{Random walk}
        \end{subfigure} \\
    \end{tabular}
    \caption{Comparison of each of the seven sub-sampling algorithms.}
    \label{fig:graph}
\end{figure}

Figure \ref{fig:graph} displays a diagram comparing the different sub-sampling algorithms.\footnote{We are greatly indebted to one of the anonymous reviewers for providing this figure.} In pane (a), we plot a small graph with $n=12$ nodes and $m=15$ edges. In (b) and (c), randomly sampled nodes are marked in orange with a black border, with their connecting edges also in black. Sampled edges are in orange while the incident nodes are colored in black with an orange border in (d). The first node of the search is represented with a white circle in the center for (e) and (f), with the other visited nodes colored in orange. The first node has the same white mark in (g) and (h). This figure reveals some insights into the similarities and differences in graph topology of the sub-graphs induced by the different sampling algorithms. 

\section*{Theoretical proofs}

\paragraph{Theorem 3.2:} {\it Given some graph $G$ with adjacency matrix $\bA$ and sampling scheme $\mathcal S$, let $S$ be a randomly chosen sub-graph. Let $\hat{\bc}$ be estimated CP labels using the CP divide-and-conquer algorithm. Then the expected mis-clustering rate, $\mathbb E\delta(\hat\bc,\bc^*)$ can be bounded by}
\begin{equation}\label{eq:theo_cp}
    \mathbb E \delta(\hat\bc,\bc^*)
    \leq \frac1n\left(\mathbb E||\hat{\bc}^{(S)}-(\bc^*)^{(S)}||_2^2+\mathbb E\sum_{i=1}^n y_i^{(S)}\right)
\end{equation}
{\it where $\hat{\bc}^{(S)},(\bc^*)^{(S)}\in\{0,1\}^{qn}$ are the estimated and true CP labels, respectively, restricted to sub-graph $S$, and $y_i^{(S)}=1$ if node $i$ is in the core and was not sampled in sub-graph $S$, and 0 otherwise.}

 \noindent
{\it Proof.} Let $\bc^*$ be the true CP labels. Let $\hat{\bc}$ be the estimated labels from the divide-and-conquer procedure where
$$
    \hat c_i = \frac1B\sum_{b=1}^B \hat c_i^{(b)}
$$
where $\hat\bc^{(b)}$ is the CP labels returned from the $b$th sub-sample. Note that $\hat c_i^{(b)}$ is 1 if the node was sampled and assigned to the core and 0 otherwise. Now, notice that
$$
    \hat c_i-c_i^*
    =\left(\frac1B\sum_{b=1}^B\hat c_i^{(b)}\right)-c_i^*
    =\frac1B\sum_{b=1}^B(\hat c_i^{(b)} - c_i^*).
$$
Thus,
$$
    \delta(\hat\bc,\bc^*)
    =\frac1n\sum_{i=1}^n\left\{\frac1B\sum_{b=1}^B(\hat c_i^{(b)} - c_i^*)\right\}^2
$$
By Cauchy-Schwarz (or Jensen's) inequality, we have
$$
    \delta(\hat\bc,\bc^*)
    \leq \frac1{nB}\sum_{i=1}^n\sum_{b=1}^B\{\hat c_i^{(b)} - c_i^*\}^2
    = \frac1{nB}\sum_{b=1}^B\sum_{i=1}^n\{\hat c_i^{(b)} - c_i^*\}^2
$$
Now, 
$$
    \frac1n\sum_{i=1}^n\{\hat c_i^{(b)} - c_i^*\}^2
    =\frac1n\sum_{i=1}^n \mathbb I(\hat c_i^{(b)}\neq c_i^*)
$$
is the mis-classification error of the algorithm on a single sub-graph. We can get an error here from one of two ways: if our sub-graph contains node $i$ but incorrectly classifies it, or if we do not sample node $i$ and it is in the core. Thus,
$$
    \frac1n\sum_{i=1}^n\{\hat c_i^{(b)} - c_i^*\}^2
    =\frac1n||\hat \bc^{(b)} - \bc^{*(b)}||_2^2 + \frac1n\sum_{i=1}^n y_i^{(b)}   
$$
where $\bc^{*(b)}$ is the true core labels restricted to sub-graph $b$ and $y_i^{(b)}=1$ if node $i$ is in the core and was not sampled in $b$, and 0 otherwise. Thus, 
$$
    \delta(\hat{\bc},\bc^*)
    \leq \frac1{nB}\sum_{b=1}^B \left(||\hat \bc^{(b)} - \bc^{*(b)}||_2^2 + \sum_{i=1}^n y_i^{(b)} \right)
$$
Since each term in the sum is independent and identically distributed, we find
\begin{equation}
    \mathbb E\delta(\hat{\bc},\bc^*)
    \leq \frac1n \left(\mathbb E ||\hat \bc^{(S)} - \bc^{*(S)}||_2^2+\mathbb E\sum_{i=1}^n y_i^{(S)}\right)
\end{equation}
where $S$ is a randomly chosen sub-graph from our sub-sampling routine. This gives us the desired result. $\square$

\noindent
{\bf Corollary 3.2:} {\it Let $\bA$ be generated from a CP-SBM with $n$ nodes and $k$ core nodes. Let $ S$ be a randomly chosen sub-graph of $\bA$ of size $q_nn$  where $q_n\in(0,1)$ may depend on $n$. Furthermore, let $y_i^{(S)}$ is the event that node $i$ is a core node and not sampled in $ S$. If RN is used to sample $ S$ and $q_n=O(1)$, then}
\begin{equation}\label{eq:cp_theory_rn}
        \mathbb E\frac1n\sum_{i=1}^n y_i^{(S)}
    = (1-q)\frac{k}{n}.
\end{equation}
{\it If $S$ is sampled via DN or RE and $q_n=o(1)$, then}
\begin{equation}\label{eq:cp_theory_dn}
    \mathbb E\frac1n\sum_{i=1}^n y_i^{(S)}
    \sim \left(1-q_n\frac{p_{12}}{p_{22}}\right)\alpha_n,
\end{equation}
{\it where $\sim$ denotes the asymptotic behavior for large $n$.}\\
{\it Proof.} (Random node) Recall that $q_n=q=O(1)$ in this setting. Let $Y^{(\mathcal S)}$ be the number of core nodes drawn in sub-graph $\mathcal S$, i.e.,
$$
    Y^{(\mathcal S)}
    =k - \sum_{i=1}^n y_i^{(\mathcal S)}.
$$
Since each node is equally likely to be sampled, then $Y^{(\mathcal S)}$ has a hyper-geometric distribution with a total population of size $n$, $k$ success states in the population, and $qn$ draws, so
$$
    \mathbb E Y^{(\mathcal S)} = \frac{qnk}{n} = qk.
$$
Thus, the expected number of core nodes which are {\it not} sampled is simply
$$
    k - qk = (1-q)k
$$
Therefore,
$$
    \mathbb E\ \frac1n\sum_{i=1}^n y_i^{(\mathcal S)}
    = (1-q)\frac{k}{n}.\ \square
$$
\noindent
For the following derivations, we note that $m_n = \sum_{i<j} A_{ij}$ is highly concentrated around its mean, $\sum_{i<j} P_{ij}$, and the error bound for $m_n$ is much smaller than that of the other (appropriately scaled) random variables. Therefore, in what follows, we will ignore the randomness of $m_n$ and consider it to be approximately equal to its expectation, i.e., if the data comes from a CP-SBM, then
$$
    2m_n\approx \mathbb E 2m_n
    =k(k-1)p_{11} + 2k(n-k)p_{12} + (n-k)(n-k-1)p_{22}
    \sim n^2p_{22}.
$$
We also assume that $q_n=o(1)$, and, without loss of generality, that the first $k$ nodes are the core nodes. Furthermore, let $z_j^{( S)}\in\{0,1\}$ be the random variable such that $z_j=1$ if draw $j$ for sub-sample $ S$ yields a core node, and 0 otherwise.

\noindent
{\it Proof.} (Degree node)  
When we sample the first node, the probability that it is a core node is
$$
    \mathbb E z_1^{( S)}
    =\mathbb P(z_1^{( S)}=1)
    =\frac{\sum_{i=1}^k d_i}{2m_n}
$$
where $d_i$ is the degree of node $i$. For core node $i$,
$$
    \mathbb Ed_i
    =(k-1)p_{11}+(n-k)p_{12}
    \sim np_{12}
$$
for large $n$. Thus, we can see that
$$
    \mathbb P(z_1^{( S)}=1)
    \sim \frac kn\frac{p_{12}}{p_{22}}.
$$
Note that for any $p_{12}/p_{22}$, there exists some $N$ such that for all $n\geq N$, $\frac kn\frac{p_{12}}{p_{22}}<1$, making this a valid probability.
Now, since we are sampling without replacement, the probability of drawing a core node changes on the ensuing draws, depending on whether the previous nodes drawn were core or periphery. However, if we can show that this change in probability is small, then we can model the number of core nodes drawn with the binomial distribution. In particular, assume that we have sampled nodes $v_1,\dots,v_{qn-1}$, and note that for any draw, the expected degree, $d$, of the node which was sampled is
\begin{align*}
    \mathbb Ed
    &=\{(k-1)p_{11}+(n-k)p_{12}\}\frac{k\{(k-1)p_{11}+(n-k)p_{12}\}}{2m_n} \\
    &+ \{kkp_{12}+(n-k-1)p_{22}\}\frac{(n-k)\{kp_{12}+(n-k-1)p_{22}\}}{2m_n}\\
    &\sim n p_{22}  
\end{align*}
Thus, on the $q_nn$th draw,
\begin{align*}
    \mathbb P(z_{q_nn}^{(\mathcal S)}=1)
    &=\frac{\sum_{i=1}^k d_i-\sum_{j=1}^{q_nn-1}\mathbb I(1\leq v_j\leq k)d_{v_j}}{2m_n-\sum_{j=1}^{q_nn-1}d_{v_j}}\\
    &\sim 
    \frac{nkp_{12} - q_nn\times \frac{k}{n}\frac{p_{12}}{p_{22}}\times n p_{22}}{n^2p_{22}-q_nn(np_{22})}\\
    &\sim \frac{nkp_{12}(1-q_n)}{n^2p_{22}(1-q_n)}\\
    &\sim \frac{k}{n}\frac{p_{12}}{p_{22}}
\end{align*}
since we assumed $q_n=o(1)$. Thus, we can model, $Y^{( S)}$, the number of core nodes drawn in sub-graph $ S$, as
$$
    Y^{( S)}\sim\mathsf{Binomial}(q_nn,\tfrac{k}{n}\tfrac{p_{12}}{p_{22}}),
$$
so $\mathbb EY^{( S)}=q_nk\frac{p_{12}}{p_{22}}$ such that
$$
    \mathbb E\frac1n\sum_{i=1}^n y_i^{( S)}
    \sim \left(1-q_n\frac{p_{12}}{p_{22}}\right)\alpha_n
$$
as we hoped to show. $\square$

\noindent
{\it Proof.} (Random edge)
We follow a similar pattern for RE. Recall that we are sampling edges now, so the probability that the first edge we sample is incident to a core node is:
$$
    \mathbb P(z_1^{( S)}=1)
    =\frac{\sum_{i=1}^k d_i}{m}
    \sim \frac{knp_{12}}{\frac12n^2p_{22}}
    \sim 2\frac{k}{n}\frac{p_{12}}{p_{22}}
$$
Note that since we assume $\alpha_n=o(1)$, the probability of drawing an edge incident to two core nodes is vanishingly small. Now, notice that since we draw two nodes every time we sample an edge, we only make $\frac12q_nn$ draws. Then wimilar to RN, for the $\frac12q_nn$th draw, the probability of drawing an edge with a core node is still
$$
    \mathbb P(z_{q_nn}^{(\mathcal S)}=1)\sim 2\frac{k}{n}\frac{p_{12}}{p_{22}}
$$
Thus, we can model $Y^{(\mathcal S)}$, the number of core nodes drawn in sub-graph $\mathcal S$ as
$$
    Y^{(\mathcal S)}\sim\mathsf{Binomial}(\tfrac12q_nn, 2\tfrac{k}{n}\tfrac{p_{12}}{p_{22}})
$$
and the result follows as before. $\square$

\section*{Additional simulation results}
\begin{figure}
    \centering
    \includegraphics[width=\linewidth]{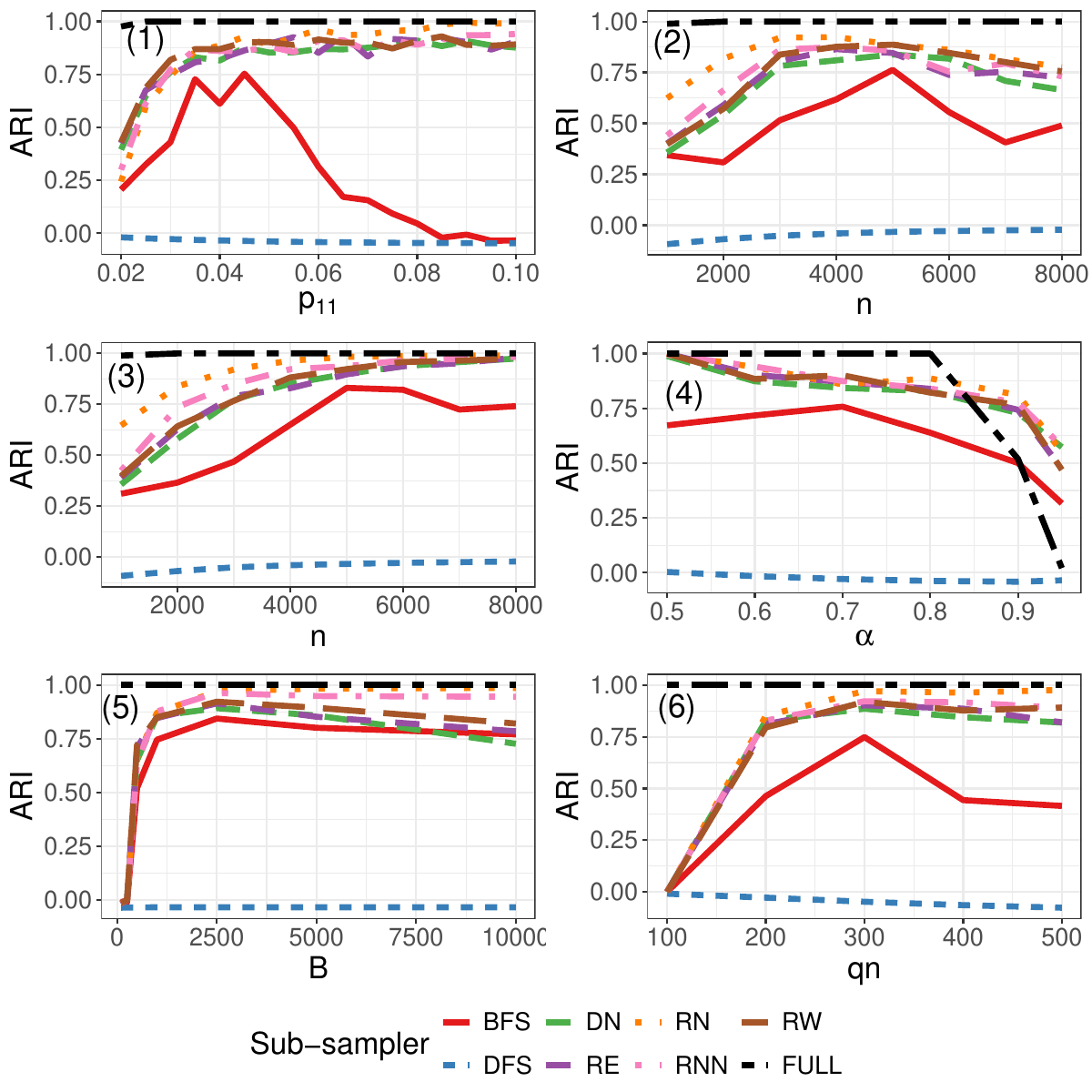}
    \caption{Community detection simulation results with Walktrap base algorithm. The number in the upper-left corner corresponds to the simulation setting (1-6).}
    \label{fig:com_resW}
\end{figure}

\begin{figure}
    \centering
    \includegraphics[width=\linewidth]{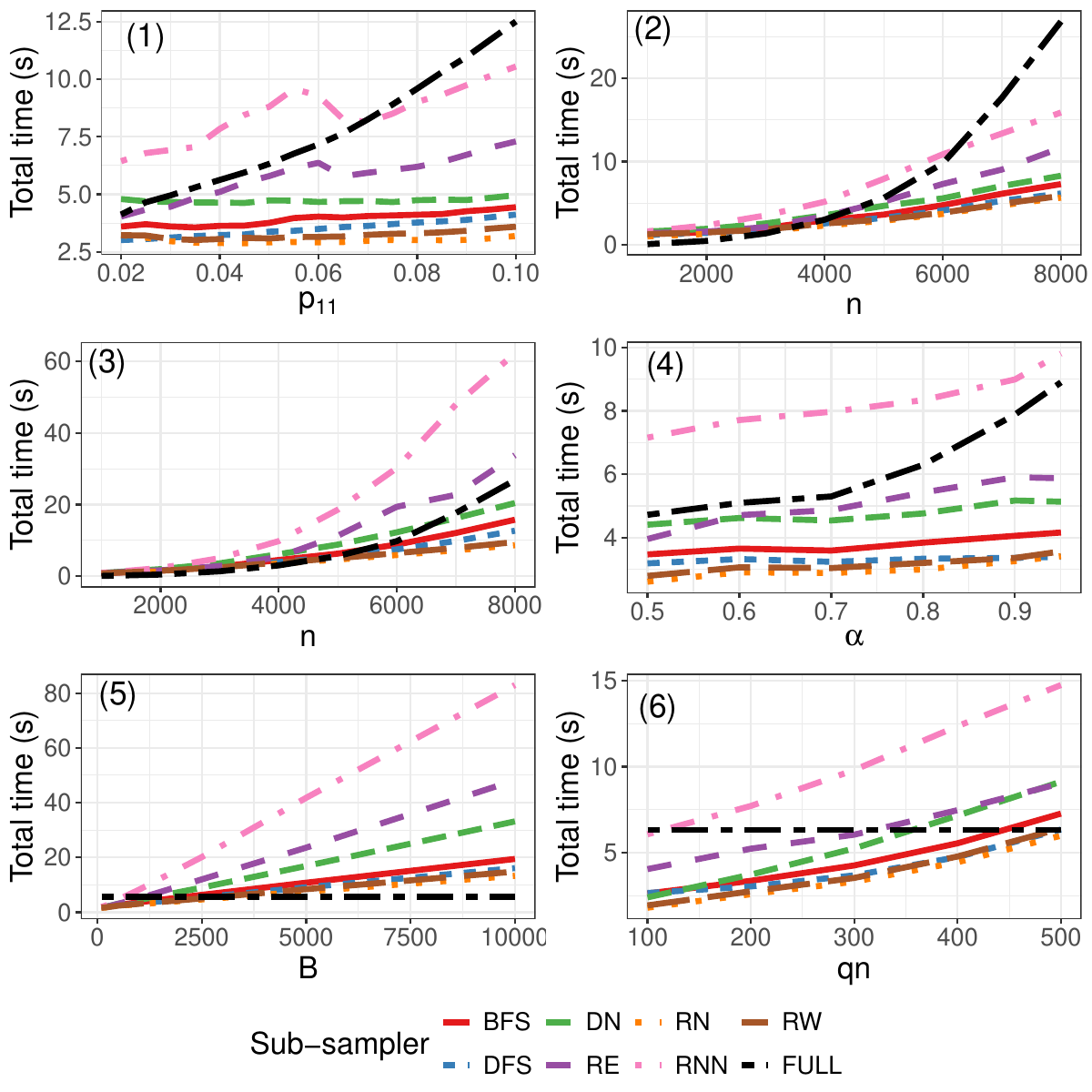}
    \caption{Community detection simulation run-time results with Walktrap base algorithm. The number in the upper-left corner corresponds to the simulation setting (1-6).}
    \label{fig:com_resTW}
\end{figure}

We replicate the simulation study from the main manuscript, now using the Walktrap community detection algorithm \citep{Pons:2005aa} as the base algorithm. We use the same settings as in the main manuscript. The results are in Figures \ref{fig:com_resW} and \ref{fig:com_resTW}. We can see that the results are very similar to those with the Fast Greedy base algorithm. The main difference is that the base algorithm outperforms $\mathsf{PACE}$ for all settings (except extremely disproportionate community sizes), but is slower for many settings. These results are more in line with what we would expect: the divide-and-conquer algorithm yields slightly worse identification results but is computationally faster.

\section*{Simulation sub-samples timing results}
\begin{figure}
    \centering
    \includegraphics[width=\linewidth]{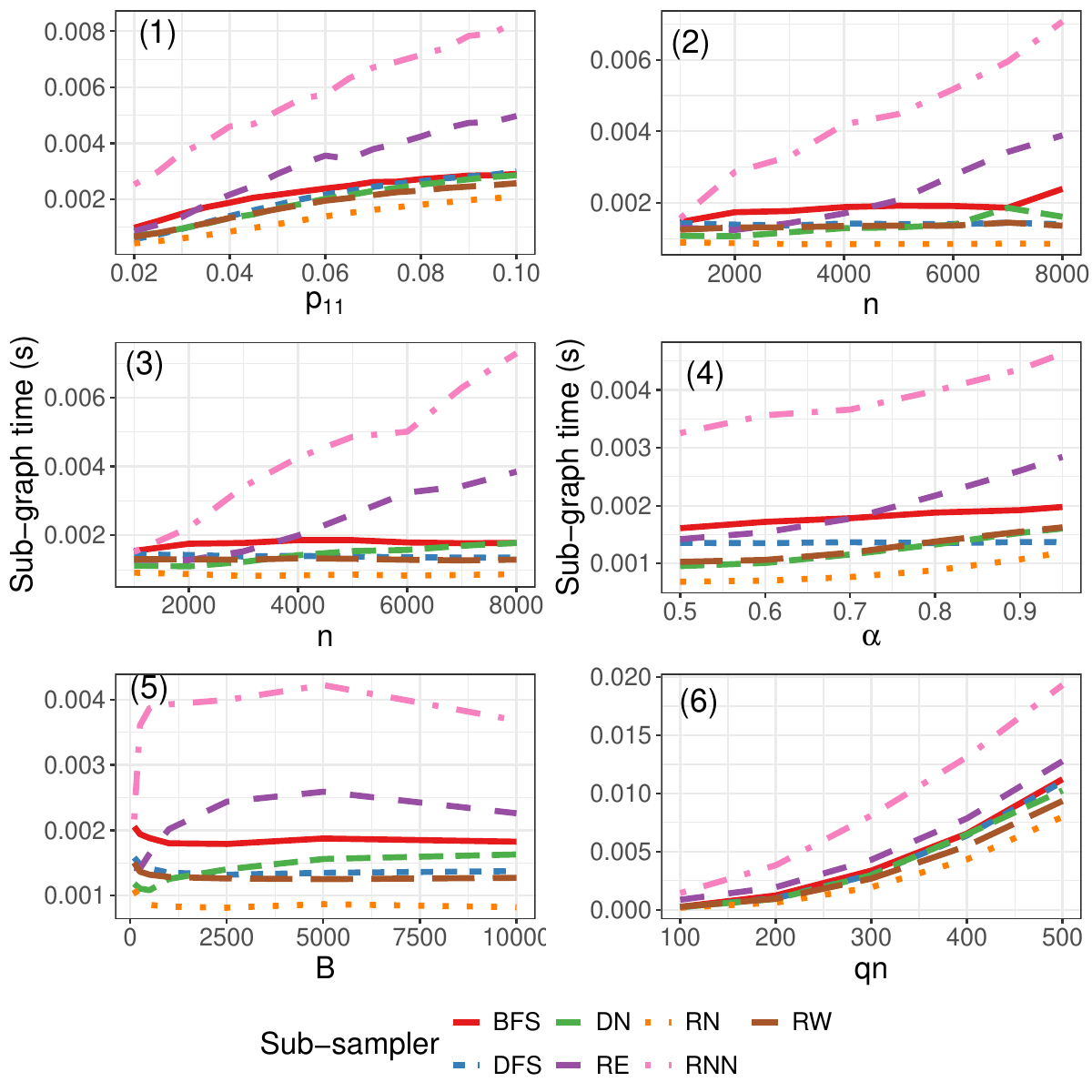}
    \caption{Average computing time to detect the community structure for each sub-sample with Fast Greedy base algorithm. The number in the upper-left corner corresponds to the simulation setting (1-6).}
    \label{fig:com_timeS}
\end{figure}

\begin{figure}
    \centering
    \includegraphics[width=\linewidth]{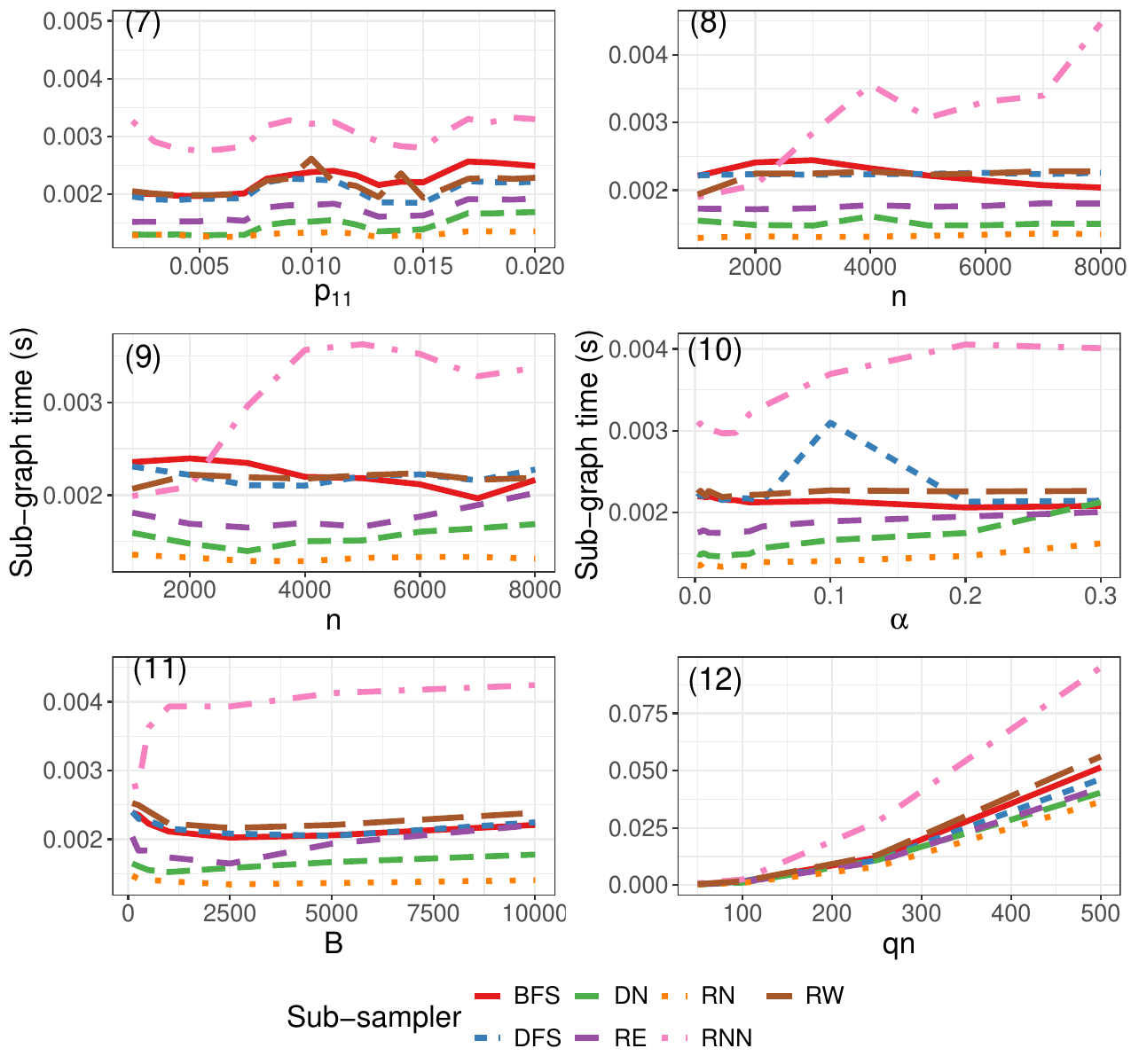}
    \caption{Average computing time to detect the CP structure for each sub-sample. The number in the upper-left corner corresponds to the simulation setting (7-12).}
    \label{fig:cp_timeS}
\end{figure}

In Figures \ref{fig:com_timeS} and \ref{fig:cp_timeS}, we report the average time it takes to detect the community structure and CP structure on the sub-samples, respectively, for the main simulation settings. In general, RNN is the slowest with RN being the fastest.

\section*{Simulation sub-samples density results}
\begin{figure}
    \centering
    \includegraphics[width=\linewidth]{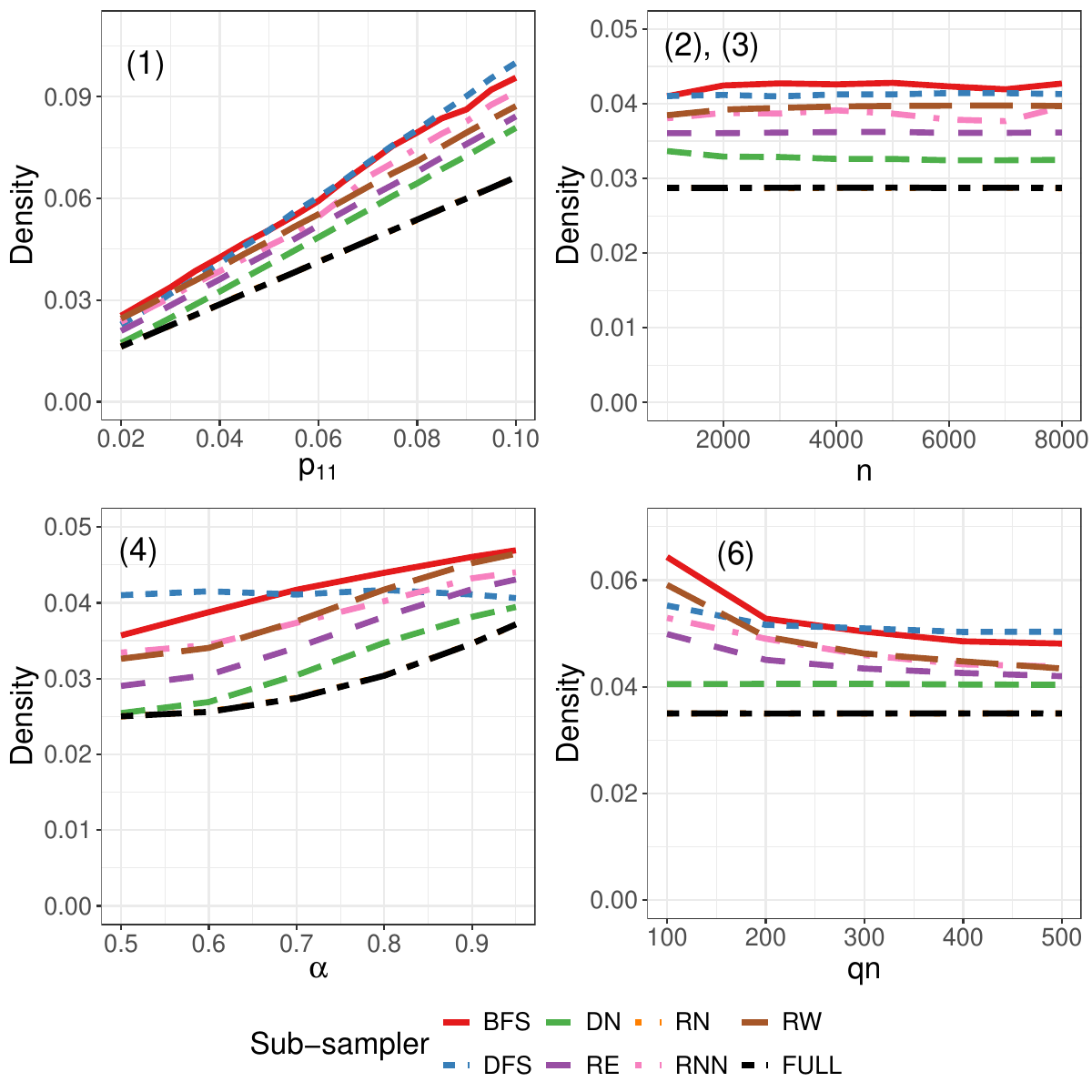}
    \caption{Average sub-graph density for community detection simulations. The number in the upper-left corner corresponds to the simulation setting (1-6). }
    \label{fig:com_dens}
\end{figure}

\begin{figure}
    \centering
    \includegraphics[width=\linewidth]{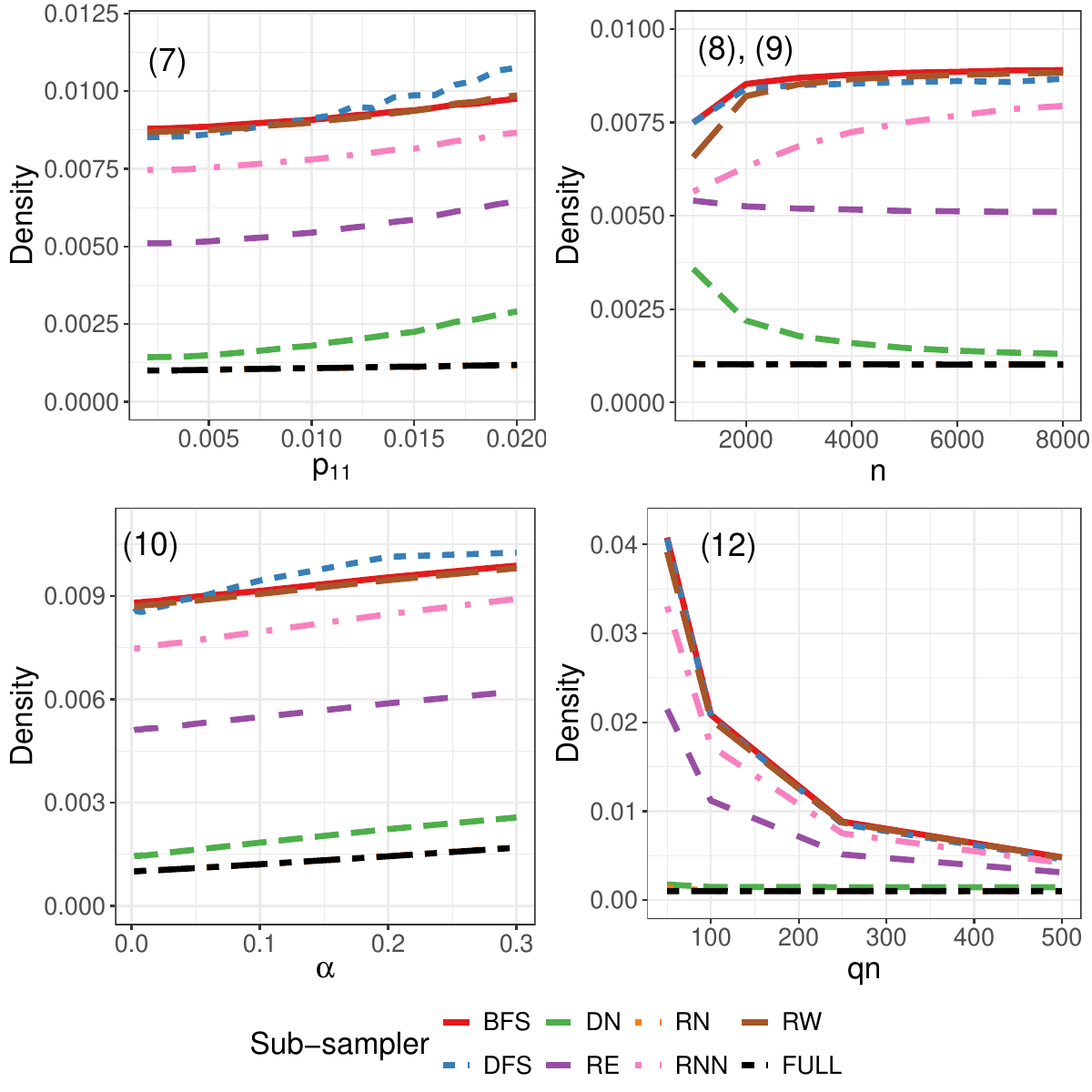}
    \caption{Average sub-graph density for CP detection simulations. The number in the upper-left corner corresponds to the simulation setting (7-12).}
    \label{fig:cp_dens}
\end{figure}

We also report the average sub-graph density for each method across each simulation setting. For each setting and parameter combination, we generate 10 networks, and then sub-sample the network 100 times for each sub-sampling algorithm. We report the average density of the sub-sample, $\hat p^{(b)}$, where
$$
    \hat p^{(b)}
   =\frac{1}{nq(nq-1)}\sum_{i,j} A^{(b)}_{ij}
$$
where ${\bA}^{(b)}$ is the adjacency matrix for sub-sampled network $\mathcal S^{(b)}$. Note that settings 3 (9) and 5 (11) are redundant for the community (CP) simulations so they are excluded. We also plot the density of the original graph. The results are in Figures \ref{fig:com_dens} and \ref{fig:cp_dens}.

\end{document}